\documentclass[pra,aps,superscriptaddress,twocolumn]{revtex4-2}

\usepackage{amsmath}
\usepackage{amssymb}
\usepackage{physics}
\usepackage{graphicx}
\usepackage{bbm}
\usepackage[colorlinks=true,linkcolor=blue,citecolor=red,urlcolor=magenta]{hyperref}
\usepackage{xcolor}
\usepackage{bbold}

\begin{document}

\title{Revealing Bell Nonlocality Does Not Require the Same Physical Measurement to be Performed Twice}

\begin{abstract}
Bell inequality violations, or Bell nonlocality for short, is not only central to testing the foundations of quantum physics but have also become an essential ingredient of numerous quantum information protocols, from quantum key distribution to the device-independent certification of randomness and secure communication.
Existing analyses of Bell nonlocality assume that a small set of predefined measurement directions can be reproduced exactly.  
But what if these directions cannot be reproduced at all? This may happen for fundamental or technical reasons, such as variations in the measurement apparatus. Here we show that the violation of a Bell inequality remains possible and that it also retains its operational significance, even when the measurement directions never repeat. 
We experimentally confirm these results using polarization-entangled photons, detecting only one particle pair per randomly chosen analyzer direction. Our results establish the robustness of Bell inequality violations against imperfect measurement repeatability and further connect Bell tests with the broader framework of randomized measurements. Moreover, they extend quantum information protocols to settings with limited measurement control,
including quantum device certification, communication complexity, and quantum cryptography.
\end{abstract}

\author{Pawe{\l} Cie{\'s}li{\'n}ski}
\thanks{These two authors contributed equally}
\affiliation{Centre for Quantum Technologies, National University of Singapore, Singapore 117543, Singapore}
\affiliation{Institute of Theoretical Physics and Astrophysics, University of Gdańsk, 80-308 Gda\'nsk, Poland}
\affiliation{Faculty of Physics, Ludwig-Maximilians University, 80799 Munich, Germany}

\author{Carlotta Versmold}
\thanks{These two authors contributed equally}
\affiliation{Faculty of Physics, Ludwig-Maximilians University, 80799 Munich, Germany}
\affiliation{Max Planck Institute for Quantum Optics, 85748 Garching, Germany}
\affiliation{Munich Center for Quantum Science and Technology, 80799  Munich, Germany}

\author{Jan W{\'o}jcik}
\affiliation{Institute of Theoretical Physics and Astrophysics, University of Gdańsk, 80-308 Gda\'nsk, Poland}

\author{Eun Mi Kim}
\affiliation{Faculty of Physics, Ludwig-Maximilians University, 80799 Munich, Germany}
\affiliation{Max Planck Institute for Quantum Optics, 85748 Garching, Germany}
\affiliation{Munich Center for Quantum Science and Technology, 80799  Munich, Germany}

\author{Jan Dziewior}
\affiliation{Faculty of Physics, Ludwig-Maximilians University, 80799 Munich, Germany}
\affiliation{Max Planck Institute for Quantum Optics, 85748 Garching, Germany}
\affiliation{Munich Center for Quantum Science and Technology, 80799  Munich, Germany}

\author{Pawe{\l} Kurzy{\'n}ski}
\affiliation{Institute of Spintronics and Quantum Information, Faculty of Physics, Adam Mickiewicz University, 61-614 Pozna\'n, Poland}

\author{Tomasz Paterek}
\affiliation{Institute of Theoretical Physics and Astrophysics, University of Gdańsk, 80-308 Gda\'nsk, Poland}
\affiliation{School of Mathematics and Physics, Xiamen University Malaysia, Sepang 43900, Malaysia}
\affiliation{Centre for Photonics and Quantum Communication Technology, Indian Institute of Technology Roorkee, Roorkee 247667, India}

\author{Harald Weinfurter}
\affiliation{Faculty of Physics, Ludwig-Maximilians University, 80799 Munich, Germany}
\affiliation{Max Planck Institute for Quantum Optics, 85748 Garching, Germany}
\affiliation{Munich Center for Quantum Science and Technology, 80799  Munich, Germany}

\author{Wies{\l}aw Laskowski}
\affiliation{Institute of Theoretical Physics and Astrophysics, University of Gdańsk, 80-308 Gda\'nsk, Poland}

\maketitle

\section*{Introduction}

Bell inequalities characterize limits on correlations between measurement results in local hidden-variable (LHV) models~\cite{Brunner2014}.
Such models generalize classical relativistic theories and are famously incompatible with quantum mechanical predictions~\cite{Bell1964}.
This very feature of quantum theory today constitutes one of the most important resources in the field of quantum technologies and is crucial in device-independent schemes for quantum key distribution~\cite{Ekert1991, Mayers_1998, Barrett_2005, Acin_2007, Pironio_2009}, randomness generation~\cite{Pironio_2010,Colbeck_2011,Colbeck_2011_2, ArnonFriedman_2018}, and communication~\cite{Cleve_1997,Brukner_2002, Brukner_2004,Moreno_2020}, among others~\cite{Brunner2014}.
The fact that Bell inequalities are violated in Nature indeed was manyfold confirmed in experiments now also closing so-called loopholes, which concern experimental imperfections that could be utilized by a LHV strategy~\cite{Hensen2015,Giustina2015,Shalm2015,Rosenfeld2017,Rauch2018, Li_2018, Storz_2023}.
Current protocols for quantum information and analyses of Bell experiments presuppose specific idealizations of experimental devices.
Can properties of real apparatus, such as variations in the implemented measurement configurations, ruin Bell’s argument?
To what extent does limited control prevent these devices from being used in applications?
Here we focus on yet unexplored imperfections related to the imprecision of switching between experimental configurations and incomplete information about the configurations actually implemented.

In Bell scenarios, each of at least two distant observers locally chooses at random between two or more measurement settings in every experimental run.
The correlations between their results make the incompatibility between LHV theories and Nature visible.
In all previous studies on Bell violation, it is assumed that the apparatus can perfectly reproduce previously chosen arrangements when set again to the same setting.
Evidently, there are limits to such ideal reproducibility, both fundamental and technical.
In reality, a particular choice of setting is in fact represented by a whole set of experimental arrangements
which differ in different experimental runs. 
For example, a polarizer never returns to exactly the same angle as set in the past, rather, one deals with a distribution of angles characterized by the precision of the experimenter or, for example, of the rotating motors.
This raises the question whether this lack of reproducibility opens a new loophole or how much of the non-repeatability we can tolerate to still violate a Bell inequality.

A natural method to answer this question is provided via randomized measurement techniques. This framework has already shown its usefulness in quantum information processing tasks~\cite{Elben2022, Cieliski2024}.
In particular, randomized measurements play a central role in shadow tomography~\cite{Huang2020, Vermersch2024}, estimation of non-linear functions of a quantum state~\cite{Brun2004, vanEnk2012, Knips2020},  measurement of many-body topological invariants~\cite{Elben2020topo}, as well as in detection and characterization of quantum entanglement~\cite{Tran2015, Tran2016, Brydges2019, Ketterer2019, Wyderka2020, Zhou2020, Elben2020, Yu2021, Imai_2021, Neven2021, Ketterer_2022}.
In the context of Bell's theorem, the probability of Bell violation under randomized measurement directions was examined thoroughly~\cite{Liang_2010,Shadbolt_2012,Wallman_2012,Fonseca_2015,Senel_2015,de_Rosier_2017,Lipinska_2018,Barasinski_2019,Barasinski_2020,Yang_2020,de_Rosier_2020,Pandit_2022,Barasinski_2021,Cieslinski_2025}, and it was shown that a sufficiently large number of local measurement outcomes or measurement settings leads to an almost certain disagreement with LHV theories~\cite{de_Rosier_2017, Lipinska_2018}.
However, all such studies assumed the exact reproducibility of the experimental arrangements corresponding to the settings and/or the availability of many state copies per measurement arrangement. 

Here we show the possibility of Bell violations using only a single state copy for each of the non-repeatable measurements.
We prove that while the quantum correlations are reduced, the correlations attainable by LHV models will never enlarge.
In the context of qubits, we represent an experimental arrangement as an effective \emph{direction} on the Bloch sphere which determines the axis on which the local projection basis lies. 
For a given setting, these directions fluctuate around the ideal value, here distributed either uniformly on a ``patch'' on the Bloch sphere or according to a quasi-normal distribution.
These distributions are natural quantifiers of the experimenter’s knowledge or ability to control the reference frame.
We derive the optimal shape of uniform patches that maximize correlation values, given a certain patch size.
The obtained Bell bounds remain intact, and the fluctuations of the measurement directions in fact can be surprisingly large. 
The methods developed here even allow to analyze the case in which measurement directions are distributed uniformly over the entire Bloch sphere. Then, quantum nonlocality can still be proven if an experimenter has at least partial knowledge about their coordinates.
For this scenario, we propose a scheme to partition Bloch spheres into regular spherical polyhedra, with facets corresponding to non-repeating directions for each setting, and construct a multi-setting Bell inequality composed of 1728 Clauser-Horne-Shimony-Holt (CHSH) expressions~\cite{Clauser1969}, which utilizes all experimental data, and where all constituting Bell expressions can be violated simultaneously.

Our theoretical findings are accompanied by quantum-optical experiments in which $10^4-10^5$ pairs of polarization-entangled photons are measured along locally random directions.
The considered types of randomness include the Haar measure for the whole Bloch sphere, uniformly random directions for patches shaped as spherical caps, as well as the von Mises-Fisher distribution (quasi-normal distribution on a sphere).
The directions are different in each and every run, and only one pair of photons is registered per one choice of directions.
Yet, we observe statistically significant Bell violations in agreement with the theory.

\section*{Results}
\subsection*{Bell scenario}

Consider two observers, Alice and Bob, operating measuring devices in distant laboratories.
In each experimental run, Alice chooses on her apparatus one of two settings denoted as $k = 1,2$, similarly Bob chooses between $l = 1,2$, and subsequently they record the outcomes $A=\pm1$ and $B=\pm1$, respectively. 
They conduct such measurements on many pairs of particles, in each experimental run again randomly selecting $k$ and $l$.
The randomness ensures that the choice of settings is independent from the hidden variables, a condition known as ``free choice''.
After a sufficient number of runs, they meet and compute the correlation functions $E_{kl} = \langle AB \rangle_{kl}$, defined as mean values of the product of measurement results obtained for the four different combinations of settings $k$ and $l$. 
Finally, they calculate the Bell-CHSH parameter~\cite{Clauser1969}:
\begin{equation}
\mathcal{B} = E_{11} + E_{12} + E_{21} - E_{22}.
\label{EQ_CHSH}
\end{equation}
This combination of correlation functions is bounded in all LHV models as $| \mathcal{B}_{\mathrm{LHV}}| \le 2$, while on the contrary, quantum mechanics predicts $|\mathcal{B}_{\mathrm{QM}} | \le 2 \sqrt{2}$~\cite{Clauser1969}.
This is achieved, for example, by a maximally entangled state $|\phi^+\rangle=(|00\rangle+ |11\rangle)/\sqrt{2}$, and settings $k$ and $l$ corresponding to the following directions on the Bloch sphere: $\vec a_1 = \hat x$, $\vec a_2 = \hat z$, $\vec b_1 = (\hat x + \hat z)/\sqrt{2}$, and $\vec b_2 = (\hat x - \hat z)/\sqrt{2}$.

\subsection*{LHV bounds and non-repeatability}

In the example above, as in all typical Bell scenarios, it is assumed that Alice and Bob can reliably switch between their two measurement directions, an assumption denoted here as \textit{repeatability}.
Let us now relax this assumption and allow many measurement directions (Bloch vectors) to be associated with a given setting. 
Operationally, when Alice chooses setting $k$, her apparatus sets one of the measurement directions $\vec a$, distributed according to a probability density $\alpha_k(\vec{a})$, possibly supported on the entire Bloch sphere.
Since the selection of $\vec a$ is a local process, Alice's measurement results for setting $k$ can be expressed as averages over a deterministic LHV model $A(\vec a, k, \lambda)$ assigning $\pm 1$ outcomes to all points on the Bloch sphere based on setting $k$ and hidden variables $\lambda$. 
The same applies to Bob.
Notably, this also holds true if the distributions of measurement directions themselves depend on the hidden variable $\lambda$, as long as the choice of settings $k$ and $l$ remains ``free''.
Therefore, the LHV bounds remain intact, and a violation of the Bell inequality observed without repeatability of measurement directions still excludes LHV models.
The lack of repeatability just represents a particular instance of a local effect (such as experimental imperfections, lack of control, or randomness) which can never \textit{increase} the power of LHV models to generate Bell-correlations but can only \textit{decrease} the ability of quantum systems to achieve them.
In the next section, the guiding question is thus: how much repeatability is needed to actually detect a violation of a Bell inequality with an entangled quantum system?

\subsection*{Quantification of non-repeatability and Bell inequality violation}

We model a realistic experimental scenario, where the lack of repeatability is independent for each observer and encoded in distributions of local measurement directions 
$\alpha_k(\vec a)$ for Alice and $\beta_l(\vec b)$ for Bob.
Recall that the quantum correlation function for a bipartite state $\rho$, measured along Bloch vector $\vec a$ and $\vec b$, is given by
\begin{equation}
E(\vec a, \vec b) = \mathrm{Tr}[ \rho (\vec a \cdot \vec \sigma \otimes \vec b \cdot \vec \sigma)],
\end{equation}
where $\vec \sigma = (\sigma_x,\sigma_y,\sigma_z)$ is the vector of Pauli matrices.
If the measurement directions of Alice and Bob vary and Alice and Bob have no further knowledge about the resulting direction beyond the chosen setting, the quantum mechanical predictions for correlation functions entering the Bell-CHSH parameter need to be averaged:
\begin{eqnarray}
E_{kl} & = & \int d \vec a \, d \vec b \, \alpha_k(\vec a) \beta_l(\vec b) \mathrm{Tr}[ \rho (\vec a \cdot \vec \sigma \otimes \vec b \cdot \vec \sigma)] \\
& = & \mathrm{Tr}[ \rho (\vec \alpha_k \cdot \vec \sigma \otimes \vec \beta_l \cdot \vec \sigma)],\nonumber
\end{eqnarray}
where $d\vec{a}$ and $d\vec{b}$ are infinitesimal area elements (where the vector arrow indicates dependence on multiple coordinates) on the Bloch spheres of Alice and Bob, respectively.
The last equality is a consequence of the linear dependence of the correlation function on the Bloch vectors $\Vec{a}$ and $\Vec{b}$, and we introduced averaged directions:
\begin{equation}
\vec \alpha_k = \int d \vec a  \, \alpha_k(\vec a) \, \vec a,
\qquad
\vec \beta_l = \int d \vec b  \, \beta_l(\vec b) \, \vec{b}.
\end{equation}
The averaging procedure is equivalent to replacing the normalized Bloch vectors by their averages, i.e., an ideal von Neumann measurement is replaced with a POVM, or more specifically, an unsharp observable (in general $|\vec{\alpha}_k|, |\vec{\beta}_l| < 1$).
Note that the shrinking factors can effectively include experimental noise and other imperfections.
The same approach generalizes to multipartite correlations.

\begin{figure}[!t]
\includegraphics[width=0.45\textwidth]{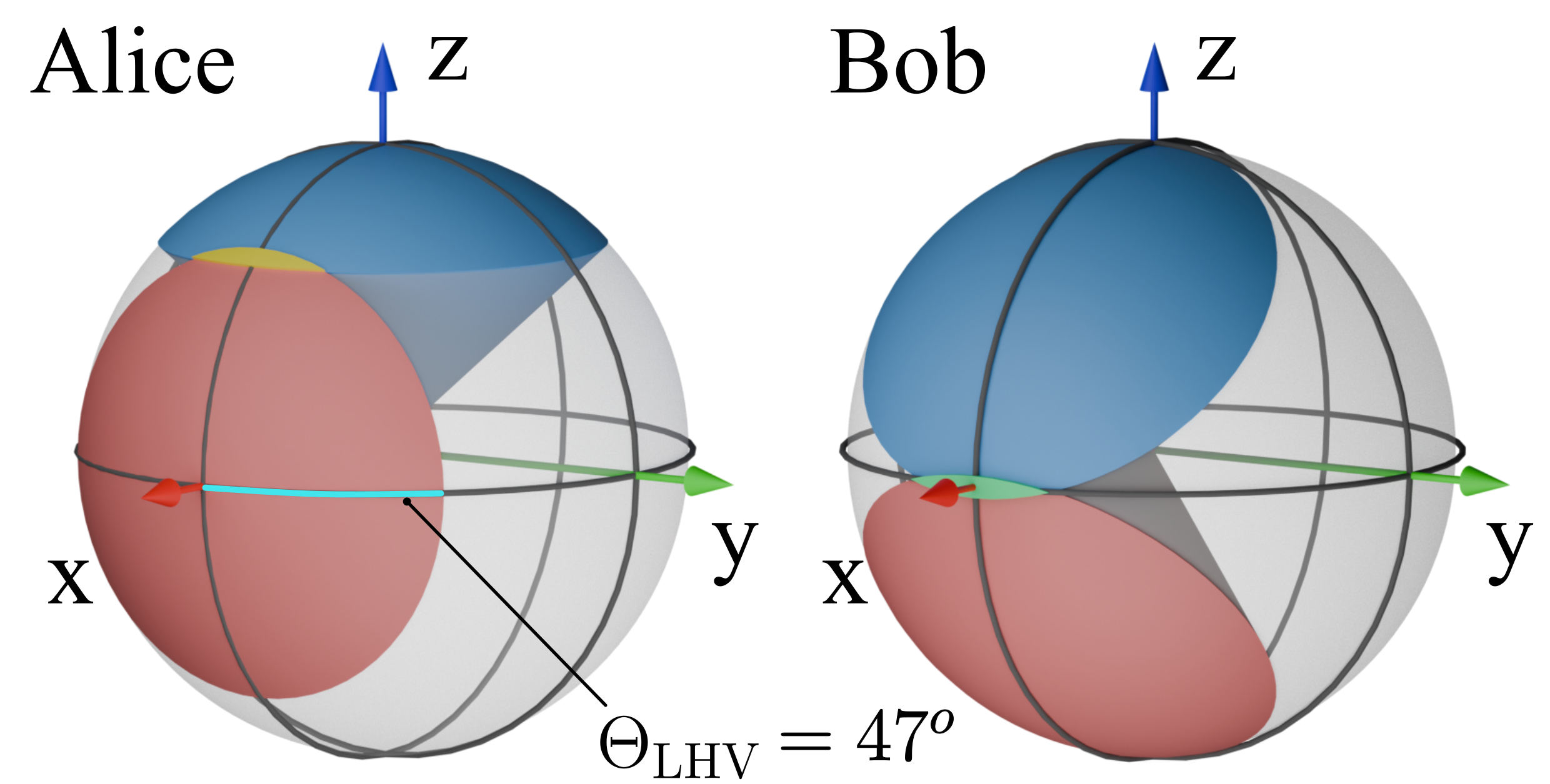}
\caption{\label{FIG_M7} Visualization of non-repeatability patches.
The two spheres are the Bloch spheres of measurement directions selected by Alice and Bob. The spherical cap (so-called ''patch'') denotes the set of directions corresponding to a given setting in the Bell parameter.
The patches are centered at their optimal directions for the violation of the CHSH inequality on a maximally entangled state.
The angle $\Theta_\mathrm{LHV}$ gives the maximal angular size of the patch still allowing observation of Bell violation.
Note that the patches overlap, i.e., some directions are not uniquely assigned to a setting.
}
\end{figure}

A natural way to capture the non-repeatability in this model is via \emph{patches} on the Bloch sphere, which are supports of uniformly distributed local measurement directions for each setting.
The patch shape that maximizes quantum correlations for a fixed area is a spherical cap, for a proof, see Methods \textit{Spherical cap as the optimal patch}.
By symmetry, the average Bloch vector of Alice's $k$th setting becomes $\vec \alpha_k = \mu \, \vec a_k$, with
\begin{eqnarray}
    \mu  &=& \frac{1}{4 \pi \sin^2(\Theta/2)} \int_0^{2\pi} d\phi \int_0^{\Theta} d \theta \sin \theta \cos \theta \nonumber \\ 
     &=&  \frac{1}{4} \frac{\sin^2 \Theta}{\sin^2(\Theta/2)}=\cos^2\frac{\Theta}{2},\label{eq:spherical_cap_length}
\end{eqnarray}
where $\theta$ is the polar angle measured from $\vec a_k$, $\phi$ is the corresponding azimuthal angle, and $\Theta$ is the polar width of the spherical cap with area $4 \pi \sin^2(\Theta/2)$.
Since $\mu \leq 1$, it acts as a shrinking factor (also referred to as the sharpness parameter).
For Bob, $\vec \beta_l = \mu \, \vec b_l$ assuming the same non-repeatability area, see Fig.~\ref{FIG_M7} for visual representation.
Note that this reasoning can be applied to any problem involving the estimation of quantum expectation values under non-repeatability conditions, e.g., in entanglement witnessing experiments.

We now ask how large $\Theta$ can be to still observe a violation of the CHSH inequality with a maximally entangled state $|\phi^+\rangle$.
The CHSH parameter in Eq.~(\ref{EQ_CHSH}), averaged over the non-repeating measurement directions, $\bar{\mathcal{B}}$, is related to the parameter $\mathcal{B}$ as $\bar{\mathcal{B}} = \mu^2 \mathcal{B}$, with $\mu^2$ acting similarly to a visibility in the Werner state.
In general, for $k$-partite correlations, the prefactor becomes $\mu^k$.
The CHSH inequality is violated for $\mu^2 > 1/\sqrt{2} \approx 0.707$, which corresponds to a maximal polar width of
\begin{equation}
    \Theta_{\mathrm{LHV}} = 2 \arccos(\sqrt{\mu}) \approx 47^{\circ}.
\end{equation}
Since the optimal Bell-CHSH measurement directions are orthogonal on the Bloch sphere, the corresponding patches overlap, see Fig.~\ref{FIG_M7}, indicating that some directions are possible for both settings.
In the case where one of the parties can perfectly reproduce the measurement directions, the non-reproducibility patch of the other observer can be as large as $\Theta_{\mathrm{LHV}}^{\mathrm{asym}} =2 \arccos{(1/2^{1/4})}\approx  65^{\circ}$. 
It is worth adding that the angle $\Theta_{\mathrm{LHV}}$ corresponds to the critical shrinking factor below which the two local POVM measurements become simultaneously measurable~\cite{Busch1986}, and hence an explicit LHV model for the generated data exist~\cite{Fine1982}.
The experimental demonstration of the discussed patch approach to non-repeatability and verification of its predictions is discussed in the \textit{Experiment} section.

While uniform patches provide a natural representation of lack of knowledge, one may also consider different probability distributions,
such as the von Mises–Fisher distribution, which is an analogue of the normal distribution on the sphere.
As it turns out, for distributions which allow a Bell violation, such as uniform patches with $\Theta < 47^\circ$, the achieved shrinking factor $\mu$ is mostly independent of the distribution's shape and approximately determined by its standard deviation.
We investigate the impact of the shape of the distribution experimentally and show the approximate independence of the shrinking factor in the Methods \textit{Modeling of non-repeatability with different
probability distributions}.

\subsection*{Uniform measurement directions}

In some experiments, most notably those concerned with entanglement detection~\cite{Brydges2019,Knips2020}, a different scenario is encountered,
where measurement directions are distributed uniformly over the entire Bloch sphere.
We therefore ask under which conditions LHV models can be disqualified in this scenario.
For uniformly distributed measurement directions, we can again use the patch approach from the previous section as a criterion to accept the measurement result only if the direction lies in a certain acceptance patch for a given setting chosen by Alice or Bob.
This adds a mechanism of loss to our scenario if the direction falls outside of the acceptance patch.
In fact, most of the data is discarded.
This can be exploited by an LHV model (for example, it can move the direction out of the patch for an unfavorable outcome assignment), and thus opens a fair-sampling loophole~\cite{ClauserShimonyReview, Larsson2014}, analogous to the detection loophole. 
Therefore, this approach does not conclusively invalidate LHV theories.
However, if \emph{this power is removed from the LHV}, which is a perfectly realistic assumption in many typical experimental scenarios, this loophole cannot be exploited.
Additionally, this opens up the possibility to use the randomness of directions itself to choose the settings for Alice and Bob.
Then, a strategy with acceptance patches, where those runs are kept for which the directions fall into one of two acceptance patches (for each party), becomes a valid (albeit inefficient) Bell test disqualifying a weaker variant of LHV theories with the aforementioned relaxation.
This situation involves an interesting trade‑off.
On one hand, increasing the angular size of an acceptance patch allows more data points to fall within it, since the number of accepted runs depends quadratically on the area.
On the other hand, smaller acceptance patches have larger shrinking factors, which enhance correlations and make a Bell violation easier to observe.
It turns out that violations can be observed for a range of acceptance patch sizes determined by the total number of experimental runs, as shown in Methods \textit{Trade-off between sharpness and statistical
significance in acceptance-patch approach for uniform
randomness}.

\begin{figure}
    \centering
    \includegraphics[width=0.7\linewidth]{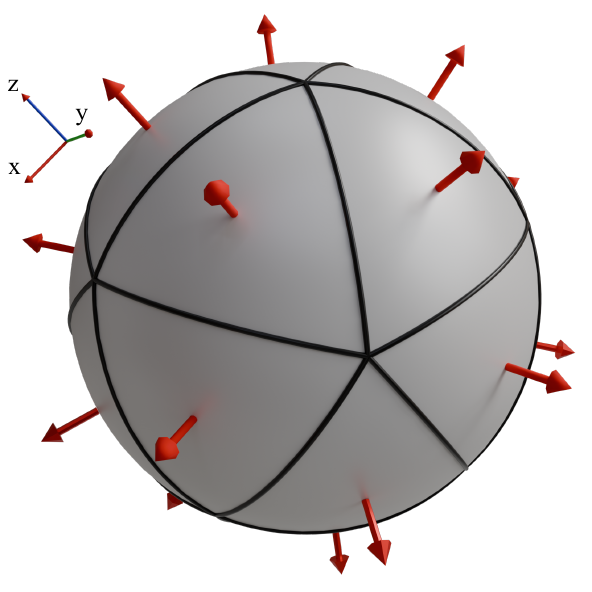}
    \caption{Icosahedral tiling of the Bloch sphere. Each spherical triangle (tile) represents the non-repeatability region associated with one local setting in a multi-setting Bell inequality. The averaged measurement directions, red arrows, coincide with the normals to the facets of a regular icosahedron inscribed in the sphere.
    }
    \label{fig:tiling}
\end{figure}

The acceptance-patch approach, as mentioned above, is inefficient as it discards most of the data. To utilize all experimental data generated with uniformly random measurement directions over the entire Bloch sphere  (still using the randomness of directions as the free choice of settings), we divide the sphere into tiles corresponding to the facets of regular spherical polyhedra and construct multi-setting Bell inequalities.
Each tile defines a non-repeatability region for one local setting and can be introduced after the whole experiment is performed.
We consider three natural choices for the tiling: spherical tilings obtained by projecting an inscribed octahedron, dodecahedron, and icosahedron onto the sphere, resulting in $8$, $12$, and $20$ tiles, per observer, respectively.
An illustration of the icosahedral tiling is given in Fig.~\ref{fig:tiling}.
By symmetry, the tile-averaged measurement direction for each setting is normal to the corresponding facet.
The lengths of the averaged Bloch vectors can be easily computed.
For the three solids, we obtain:
\begin{eqnarray}
    \mu_{8} & \approx & 0.866 , \\
    \mu_{12} & \approx & 0.916 ,\\
    \mu_{20} & \approx & 0.943 ,
\end{eqnarray}
where $\mu_n$ is the magnitude of the average direction for a tiling with $n$ facets.

As before, a bipartite CHSH violation requires $\mu^2 > 1/\sqrt{2} \approx 0.707$.
All three values listed above meet this criterion, provided only two tiles per party were used in the inequality.
Motivated by this observation, we investigate violations of multi-setting Bell inequalities built as sums of multiple CHSH terms. 
We focus on the icosahedral case, as it yields the largest average direction among the spherical tilings considered here, see~\cite{Hoi_2025, Johnson1966, zalgaller1969convex} for other examples.
An icosahedron has $20$ facets, which allows us to construct ${20 \choose 2}^2 =  36100$ distinct CHSH expressions using all available settings.
We take the general form of a single CHSH parameter 
\begin{eqnarray}
     \mathrm{CHSH}_{i,j; \, k,l}^{s_1s_2s_3s_4} &=& (-1)^{s_1} A_i B_k + (-1)^{s_2} A_j B_k \nonumber \\
     &+&(-1)^{s_3} A_i B_l+ (-1)^{s_4} A_j B_l,
\end{eqnarray}
and optimize the sum of all expressions over the choice of signs $(s_1s_2s_3s_4)$, and over the assignment of settings $i,j$ (Alice) and $k,l$ (Bob).

As a first step, we find a Bell inequality that utilizes all of the icosahedral tiles on each Bloch sphere, but not yet all the experimental data (since not all combinations of tiles are included).
The resulting twenty-setting inequality composed of ten CHSH expressions reads  
\begin{eqnarray}
    I &=& \mathrm{CHSH}_{1,12; \, 2,9}^{0100} + \mathrm{CHSH}_{2,17; \, 6,17}^{1110} + \mathrm{CHSH}_{3,4; \, 8,20}^{0111} \nonumber \\
    & + & \mathrm{CHSH}_{5,16; \, 10,14}^{0111} + \mathrm{CHSH}_{6,18; \, 1,15}^{0100} + \mathrm{CHSH}_{7,19; \, 3,18}^{1110} \nonumber \\
    & + & \mathrm{CHSH}_{8,9; \, 11,16}^{1011} + \mathrm{CHSH}_{10,14; \, 7,12}^{0100} + \mathrm{CHSH}_{11,13; \, 5,13}^{1000} \nonumber \\
    & + & \mathrm{CHSH}_{15,20; \, 4,19}^{1110} \leq 20 \, .
    \label{eq:multi_setting_CHSH}
\end{eqnarray}
As before, the inequality is optimized for the $\ket{\phi^+}$ Bell state.
The explicit ideal directions
are listed in Table~\ref{tab:settings_ico} (Methods \textit{Measurement directions for the icosahedral
tiling}), and correspond to the normalized, tile-averaged directions shown in Fig.~\ref{fig:tiling}.

In order to demonstrate how all data obtained with Haar-random measurement directions can be incorporated, we construct a Bell inequality composed of $1728$ CHSH expressions with a white noise critical visibility $v=0.781$, assuming perfect repeatability.
For completeness, the explicit inequality is provided in the arXiv source files.
%Supplementary Online Material.
The corresponding classical bound is $3456$, while a quantum value of more than $4424$ is achieved when all individual terms are violated. 
The non-repeatability results in a shrinking factor $\mu_{20}^2 = 0.889$, which diminishes the observable Bell parameter to $3932.9$ and increases the critical visibility to $v/\mu_{20}^2$.
Therefore, there is only a narrow margin left for experimental imperfections.
In the next section, we show that this violation is indeed observed experimentally. 

\subsection*{Experiment}

We performed two-qubit photonic experiments with $N_{\mathrm{tot}}= 10^4 - 10^5$ uniformly randomized measurement directions to demonstrate Bell violations with lack of measurement repeatability.
Our experimental setup is based on a type-II spontaneous parametric downconversion source in which two collinear downconversion sites are coherently overlapped \cite{Evans2010}.
With this source, we are able to produce polarization-entangled photon pairs in a $\ket{\phi^+}$ state with fidelity 0.988.
The photons are sent to two polarization analysis stages, each consisting of a quarter-wave plate, a half-wave plate, and a polarizing beam splitter.
We measure coincident detections between Alice and Bob.
The respective directions of Alice's and Bob's measurement bases were set by rotating the waveplates of the analysis stages.
A single coincidence event was recorded and used to estimate the quantum correlation functions over the non-repeatability regions.
The corresponding errors for the averaged correlation functions were computed through an unbiased variance estimator, which was propagated to the CHSH parameter uncertainty, taking into account correlated noise.

\begin{figure*}
\centering
\includegraphics[width=0.96\textwidth]{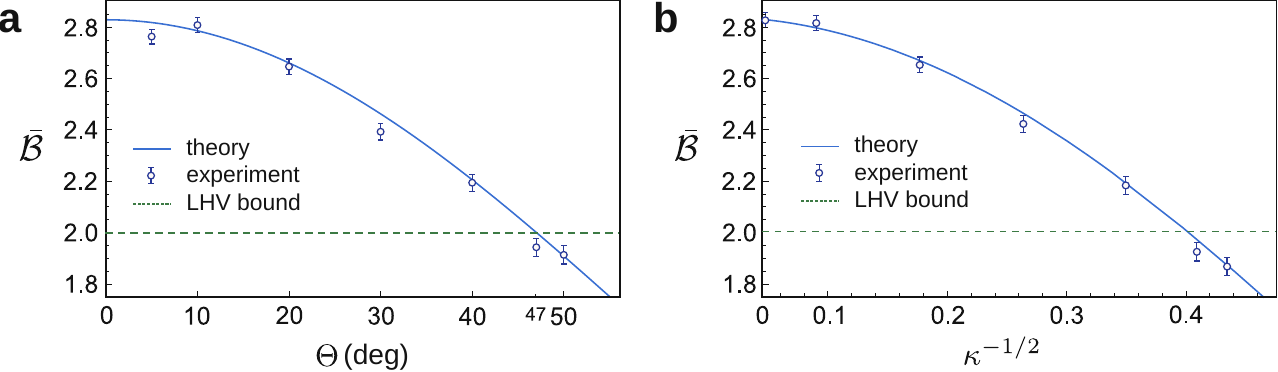}
\caption{CHSH parameter $\bar{\mathcal{B}}$ for the correlation functions averaged over the $a)$ spherical cap (uniform distribution over non-repeatability region) and $b)$ von Mises-Fisher distribution.
The solid blue curve in panel $a)$ correspond to $2\sqrt{2}\cos^4\frac{\Theta}{2}$ predicted by the patch model, see Eq.~(\ref{eq:spherical_cap_length}).
Analogous curve on panel $b)$ corresponds to the $2\sqrt{2} \mu_{\kappa}^2$, with $\mu_{\kappa}=\coth \kappa - 1/\kappa$, as a function of $\kappa^{-1/2}$.
The concentration parameter $\kappa$ controls the spread of probability over the Bloch sphere.
Blue points, together with error bars, represent the CHSH value evaluated on the experimental data with $10^4$ directions distributed $a)$ uniformly on a patch of width $\Theta$ and $b)$ according to the von Mises-Fisher distribution with concentration parameter $\kappa$, on the $|\phi^+\rangle$ Bell state in \emph{two separate experiments}.
The LHV bound in both panels is denoted by the dashed horizontal line. 
\label{FIG_EXP_PATCHES}}
\end{figure*}

For the uniform patch scenario, Alice and Bob sampled each $N_{\mathrm{tot}}=10^4$ Haar random directions within the spherical cap patches of angular widths $\Theta=5^{\circ},10^{\circ},20^{\circ},30^{\circ},40^{\circ},47^{\circ}$ and $50^{\circ}$ degrees centered at the optimal directions for evaluating the CHSH inequality on the $|\phi^+\rangle$ state.
The theory predicts that the quantum value of the CHSH parameter should scale as $\cos^4\frac{\Theta}{2}$, which is confirmed experimentally in Fig.~\ref{FIG_EXP_PATCHES}a.
All of the violations are significant.
As predicted, for $\Theta \gtrsim 47$ degrees, no violation occurs.
The same data was also used to evaluate the approach with acceptance-patches presented in Methods \textit{Trade-off between sharpness and statistical
significance in acceptance-patch approach for uniform
randomness}, where appropriate subsets of measured directions were chosen in order to simulate higher values for $N_\mathrm{tot}$.
For the measurement directions distributed according to the von Mises-Fisher distribution, we again used $N_{\mathrm{tot}}=10^4$.
The concentration parameters $\kappa$ for each distribution were chosen such that they roughly cover the same range of shrinking factors as for the uniform distributions.
The resulting quantum value of the CHSH inequality as a function of $\kappa^{-1/2}$ is shown in Fig.~\ref{FIG_EXP_PATCHES}b.
Again, the predicted scaling of $\bar{\mathcal{B}}$ is confirmed experimentally.
In particular, for $\kappa^{-1/2} \gtrsim  1/\sqrt{6.285} \approx 0.4$, the local bound is never exceeded. 
The choice of the $x$ axis scaling is motivated by the width of a highly concentrated von Mises-Fisher distribution being proportional to $\kappa^{-1/2}$, see Methods \textit{Modeling of non-repeatability with different
probability distributions} for more details. Explicit values of the Bell parameter for both families of distributions, together with the experimental errors, are presented in Table~\ref{tab:exp_patch}.

\begin{table}[ht!]
\centering
\begin{tabular}{c|c|c|c|c|c|c|c}
\hline \hline
$\Theta (\text{Deg})$ & 5 & 10 & 20 & 30 & 40 & 47 & 50 \\
\hline
$\bar{\mathcal{B}}$ & 2.76 & 2.81 & 2.65 & 2.39 & 2.19 & 1.94 & 1.91 \\ \hline
Error & 0.03 & 0.03 & 0.03 & 0.04 & 0.04 & 0.04 & 0.04 \\
  \hline \hline
$\kappa$ & 525 & 131 & 33 & 14.7 & 8.32 & 6.06 & 5.37 \\
\hline
$\bar{\mathcal{B}}$ & 2.80 & 2.79 & 2.63 & 2.41 & 2.17 & 1.92 & 1.86 \\
\hline
Error & 0.03 & 0.03 & 0.03 & 0.03 & 0.04 & 0.04 & 0.04 \\
\hline
\end{tabular}
\caption{
\label{tab:exp_patch}
Averaged CHSH parameter $\bar{\mathcal{B}}$ for spherical caps with different polar widths $\Theta$ or concentration parameters $\kappa$ from the von Mises-Fisher distribution, with corresponding errors.
}
\end{table}

As for the tiling approach, $N_{\mathrm{tot}} = 10^5$ measurement pairs were drawn from entire Bloch spheres and the registered data were grouped into the tiles of the spherical icosahedron, see Fig.~\ref{fig:tiling}.
Again, for each pair of random directions, only one coincidence event was used in data analysis.
The observed value of $I$ in Eq.~(\ref{eq:multi_setting_CHSH}) is $23.3 \pm 0.3$, clearly above the LHV bound of $20$. 
An even better violation is obtained when using all the experimental data.
We evaluated the Bell parameter combining $1728$ CHSH expressions and found that its classical bound of $3456$ was exceeded to $3941 \pm 19$, i.e., the observed violation is by $ 19.5$ standard deviations, assuming a Gaussian error distribution for hypothesis testing.
The experimental value agrees with the quantum predictions rescaled by the shrinking factor of the icosahedral tiling within at most $0.2\%$ relative error.

\section*{Discussion}

Non-repeatability appears in any experiment measuring quantum correlations: it arises not only from intentional switching between settings, as in Bell tests, but also from drifts and other experimental imperfections.
Since every correlation measurement depends on well-defined and stable reference directions, any deviation modifies the measured expectation values.
Non-repeatability is therefore an inherent constraint on all correlation-based experiments, independent of whether they are used to test LHV models.

We have proposed and tested experimentally two approaches that allow to observe Bell violation even if there is significant non-repeatability between each consecutive measurement.
A central question is whether the non-repeatability is in the realm of parameters controlled by the LHV or not.
In the former case, we showed how appropriate limits on the range of non-repeatability, i.e., distributions of actual measurement directions which are sufficiently concentrated around the ideal ones, still enable significant violations of CHSH inequalities.
Crucially, in this case, the limiting factor is not that LHV models obtain more power to generate correlations, but rather that with too little repeatability, observations on entangled systems fail to generate stronger correlations than those in LHV models.
When the randomness of directions lies beyond the control of the LHV, we developed the tiling method involving multi-setting inequalities.
It succeeds in providing significant Bell violations even if the measurement directions are distributed uniformly over the entire local Bloch spheres.

Expressing Bell experiments within the framework of randomized measurements provides a fresh perspective on entanglement and Bell nonlocality being distinct resources~\cite{Werner1989,Brunner2014,Augusiak2015,Quintino2015}.
It is well known that, even in the absence of local and shared reference frames, bipartite entanglement can be revealed from the moments of probability density characterizing the likelihood that Alice and Bob observe specific local expectation values and correlations~\cite{Tran2015, Tran2016, Brydges2019, Ketterer2019, Wyderka2020, Zhou2020, Elben2020, Yu2021, Neven2021}.
However, even complete knowledge of such a probability density cannot demonstrate Bell nonlocality. As shown in Methods \textit{No Bell nonlocality from distributions of correlations}, the statistics are compatible with an LHV model.
Moreover, even partial reference frames can be insufficient to witness Bell violations.
For instance, if each measurement direction is known only up to a horizontal plane on the Bloch sphere, i.e., only the polar angle is specified, then the quantum correlations averaged over local azimuthal angles still admit a LHV model.
In consequence, it is impossible to formulate any non-trivial Bell inequality on such a dataset, illustrating that not only complete ignorance but also certain structured forms of partial directional knowledge are insufficient for observing Bell nonlocality.
Instead, all of the approaches introduced in this work rely on partial knowledge (or even control) which allows to group the outcomes based on some measure of \textit{closeness} of directions on the Bloch sphere.

Randomized measurements have appeared in a number of applications, and we wish to point out that different applications exploit different forms of randomness.
In shadow tomography, an observable is chosen at random, but it is known which measurement direction has been selected, and it is assumed that many copies of a quantum state are available for the measurement.
The entanglement classification protocols solely use estimators obtained on many copies of the quantum state per random and unknown measurement direction.
In the present work, there is only one copy of the state per measurement direction, but many copies per measurement setting, and only partial knowledge of the measurement direction is required.

Imperfect repeatability can be interpreted more generally as a limitation on the quality of the experimenter’s reference frame.
A perfect reference frame would define Bloch-sphere directions with infinite precision, enabling exact reproducibility of any measurement setting, but such a frame is unattainable.
In practice, experimenters always operate with partial knowledge of their true measurement directions, and this uncertainty propagates to all observable quantities, influencing the certification of quantum resources.
A straightforward example is the weakening of measurement incompatibility and the increase of measurement disturbance.
Moreover, the imperfect repeatability is effectively equivalent to an unsharp measurement, which has proven itself useful in many quantum information tasks beyond the ones examined here, see, e.g.~\cite{Anwer2021, Miklin_2020, Anwer_2020, Curchod_2017, Silva_2015, Palacios-Laloy_2010, Armen_2002, Audretsch_2001}.
Our results indicate that it is fruitful to analyze the structure of the limitations on the experimenter's knowledge and control
further motivating a dedicated resource theory of reference frames and measurement control.
This motivation arises from experimental intuitions, and offers a complementary foundation to earlier approaches that developed reference frame resource theories primarily through the notions of symmetries and group invariants, see e.g.~\cite{Bartlett_2006, Vaccaro_2008, Gour_2008, Gour_2009,Barlett_2007}.
Establishing such a framework, or expanding the existing results in the discussed direction, could provide a systematic method for understanding and quantifying how imperfect directional information constrains quantum experiments.

Finally, in our approach, the non-repeatability is effectively described by a generalized measurement characterizing all directions for a given setting.
In this context, we wish to mention the connection between our approach and coarse-grained measurements~\cite{Kofler2007, Kofler2008, Teh2023}.
While coarse-graining is traditionally used in the context of macroscopic realism~\cite{Leggett1985} to describe the quantum-to-classical transition, here it provides a natural framework for quantifying the lack of repeatability in measurements on microscopic systems.

In summary, we have framed the concept of measurement non-repeatability in Bell experiments within the broader context of randomized measurements, where directions representing a single setting need not repeat, as occurs in all practical realizations of quantum measurements.
This perspective bridges a gap in the literature on certifying entanglement and Bell nonlocality using randomized measurements.
We explored the boundary of non-repeatability both theoretically and in an optical experiment, demonstrating that quantum information applications relying on Bell violations, such as communication complexity or cryptography~\cite{Brunner2014}, can be implemented without precise control over the experimental apparatus.
Our results motivate the further development of a resource-theoretic framework for reference frames in quantum experiments.

\section*{Methods}

In this section, we provide the technical details and discuss possible extensions of the findings presented in the \textit{Results} section.

\subsection*{Spherical cap as the optimal patch}
\label{APP_CAP}

In the main text, we have shown that correlations averaged over a patch are characterized by local unsharp measurements.
Here we show that, among all shapes with a fixed surface area on the Bloch sphere, the spherical cap maximizes the averaged Bloch vector representing an effective measurement and therefore yields the maximal quantum correlation.

For any shape $S$, let $\vec{\mu}$ denote the average of the measurement directions within $S$. Let us set the direction of $\vec{\mu}$ as the $z$ axis. The contribution of each $\vec{a} \in S$ to the length of $\vec{\mu}$ is given by the cosine of the angle between $\vec{a}$ and $\vec{\mu}$, denoted as $\cos \theta$, so that
\begin{equation}
    \mu = \frac{1}{\mathcal{A}}\int_S \cos \theta ds
    = \frac{1}{\mathcal{A}}\int_0^{2 \pi}\int_0^{\pi} d\phi d\theta \, f_S(\theta, \phi) \cos\theta \sin\theta , 
\end{equation}
where $\mathcal{A}$ is the area of the shape and $f_S$ is an indicator function. 
To find the shape that maximizes $\mu$ under the fixed-area constraint $\int_0^{2 \pi}\int_0^{\pi} d\phi d\theta \, f_S (\theta, \phi) \sin\theta =\mathcal{A}$, we use the Lagrange multipliers method. The corresponding Lagrangian is given as
\begin{eqnarray}
    \mathcal{L}=\frac{1}{\mathcal{A}} f_S(\theta, \phi) \cos\theta \sin\theta - \lambda ( f_S(\theta, \phi) \sin\theta - \mathcal{A}) \nonumber\\
    =\frac{1}{\mathcal{A}} f_S(\theta, \phi) (\cos\theta - \lambda \mathcal{A})\sin\theta + \lambda \mathcal{A} 
\end{eqnarray}
where $\lambda$ is the Lagrange multiplier. Since $f_S$ takes values in $\{0,1\}$ and the second term is constant, the above achieves an extremal point through its maximum when
\begin{eqnarray}
    f_S(\theta, \phi ;\lambda)
=
\begin{cases}
1, & \cos\theta>\lambda \mathcal{A},\\
0, & \cos\theta<\lambda \mathcal{A}.
\end{cases}
\end{eqnarray}
Since $\cos \theta$ is monotonous and decreasing in $[0,\pi]$, the optimal shape $S$ is given as
\begin{eqnarray}
    S=\{(\theta, \phi): \theta < \textrm{const} \},
\end{eqnarray}
where the constant is chosen such that the total area of $S$ is $\mathcal{A}$. This describes a spherical cap.

\subsection*{Modeling of non-repeatability with different probability distributions}

\begin{figure}
    \centering
    \includegraphics[width=0.7\linewidth]{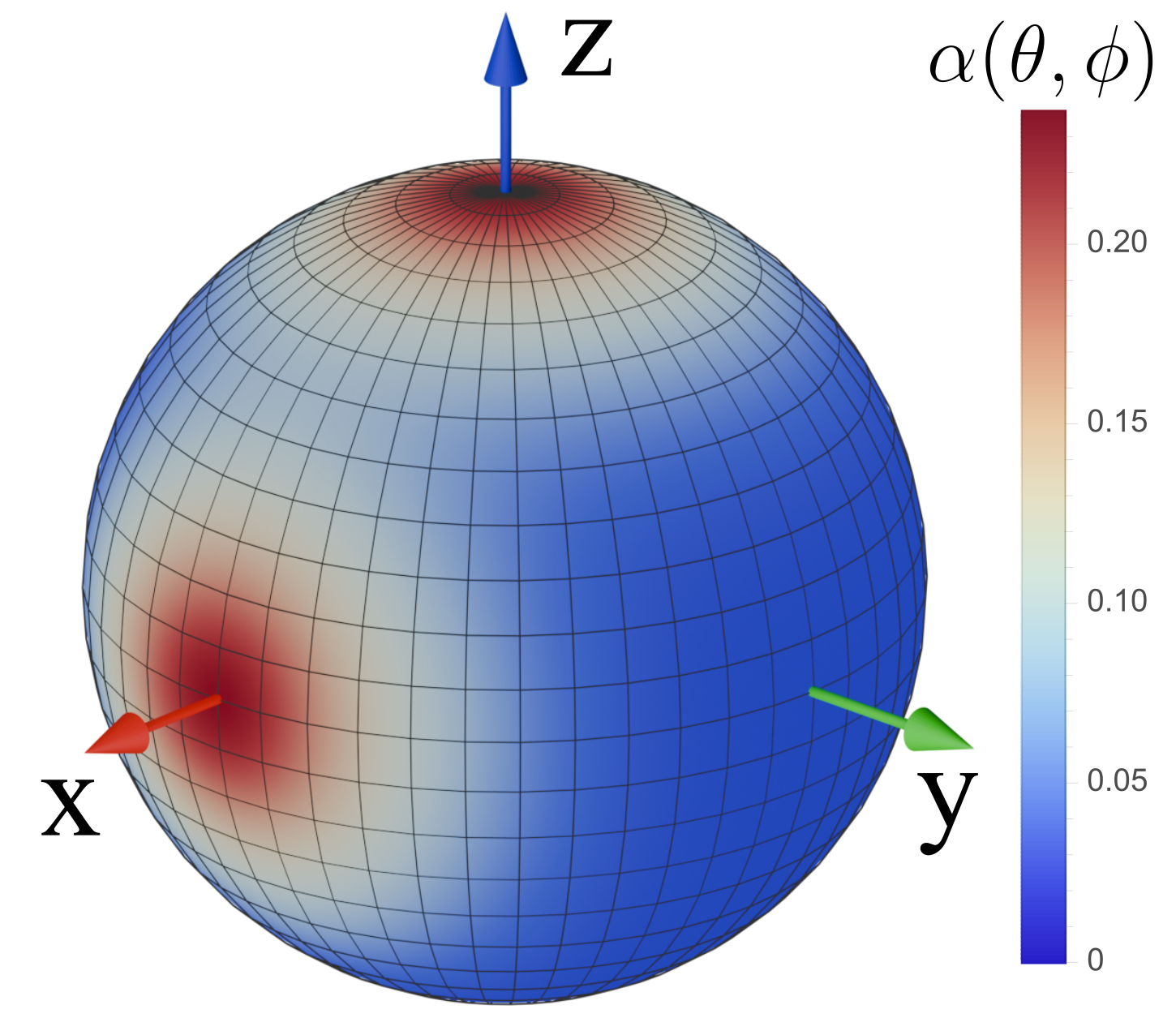}
    \caption{Von Mises–Fisher distributions on the Bloch sphere for mean directions along $z$ and $x$. The concentration parameter for this plot is $\kappa=2\pi$. While the distribution is non-zero over the entire sphere, it decays as $\sim e^{\kappa \cos\theta}$, contributing significantly only near the mean direction.}
    \label{fig:Fisher}
\end{figure}

In the \textit{Results} section, we modeled non-repeatability using patches, i.e., uniform distributions of measurement directions over a fixed area on a sphere.
To investigate how different distributions of measurement directions affect the Bell inequality, we explore the natural example of the von Mises-Fisher distributions, which is an analogue of the normal distribution on the two-dimensional sphere.
For a mean direction aligned with the north pole (always corresponding to the ideal measurement direction at hand), the probability density reads
\begin{equation}
    \alpha_{\kappa}(\theta, \phi) = \frac{\kappa e^{\kappa \cos\theta}}{4\pi \sinh \kappa } \sin \theta,
    \label{eq:vMF_dist}
\end{equation}
where $\kappa$ is a concentration parameter, and $\theta \in [0, \pi]$ is the polar angle (note that this distribution is axially symmetric).
In contrast to the patching approach, this distribution extends over the entire sphere, though it can be strongly peaked around the mean direction, as illustrated in Fig.~\ref{fig:Fisher}.
For this distribution, the length of the average Bloch vector is
\begin{equation}
    \mu_{\kappa} = \coth (\kappa) - \frac{1}{\kappa}.
\end{equation}
The minimal concentration $\kappa$ required for CHSH violation, analogous to the maximal angular width in the patch approach, is determined by $\mu_\kappa^2 = 1/\sqrt{2}$, yielding $\kappa_{\mathrm{LHV}} \approx 6.285$.
Thus, for all $\kappa > \kappa_{\mathrm{LHV}}$, CHSH violations are possible with non-repeating measurements distributed according to the von Mises–Fisher density.
For additional perspective, when $\kappa \approx \kappa_{\mathrm{LHV}}$ roughly $86\%$ of probability lies within $\theta\leq 47^{\circ}$ corresponding to the $\Theta_{\mathrm{LHV}}$ from the patch approach.

\begin{figure}
\centering
\includegraphics[width=0.48\textwidth]{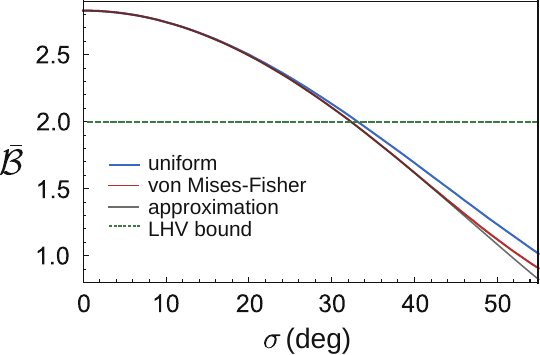}
\caption{CHSH parameter $\bar{\mathcal{B}}$ for correlation functions averaged over a uniform patch with zero mean (blue line) and von Mises-Fisher distribution (red line) as a function of standard deviation $\sigma$, together with its quadratic approximation (\ref{eq:std_approx}) plotted in black line.
Both of the examined distributions give almost the same values for the entire region of $\sigma$ where the CHSH value exceeds the local bound (dashed green line).
In other words, the CHSH inequality violation is approximately independent of the distribution used for averaging correlations and is determined solely by the value of $\sigma$. }
\label{fig:sigmaPlot}
\end{figure}

While a similar analysis can be applied to any family of distributions, it has been pointed out in the \textit{Results} section that for distributions which are sufficiently concentrated around the ideal directions, the shrinking factor $\mu$ is mostly independent of the distribution's shape and determined by the standard deviation.
To compare different families of distributions, it is necessary to find a common measure of how much the resulting measurement directions diverge from the ideal one.
For axially symmetric distributions (such as uniform patches or the von-Mises-Fisher distribution), this divergence is fully captured by the distribution of the polar angle $\theta$, $p(\theta)$, with $\theta \in [0,\pi]$.
We obtain a natural measure of the fluctuation around the pole, i.e., the ideal direction, as the standard deviation $\sigma = \sqrt{\langle \theta^2 \rangle}$ of a symmetrized distribution $p^\prime(\theta)= p(|\theta|)/2$, with $\theta \in [-\pi,\pi]$ and $\langle \theta \rangle = 0$.
This leads to a universal relationship which can be seen by expanding the expression for the shrinking factor $\mu$ as
\begin{align} \label{eq:std_approx}
    \mu  &= \int_{-\pi}^{\pi} d \theta \cos \theta p^\prime(\theta) \nonumber = \int_{-\pi}^{\pi} d \theta \big(1 - \frac{\theta^2}{2} + \mathcal{O}(\theta^4) \big) \, p^\prime(\theta) \nonumber \\
     &= 1 - \frac{\sigma^2}{2} + \int_{-\pi}^{\pi} d \theta \mathcal{O}(\theta^4)  p^\prime(\theta) \nonumber, \nonumber \\
     &= 1 - \frac{\sigma^2}{2} + \sum^{\infty}_{n=2} (-1)^n  \frac{M_{2n}}{(2n)!}, \nonumber \\
     &\approx 1 - \frac{\sigma^2}{2} \, ,
\end{align}
where the approximation holds whenever a sufficiently large mass of the distribution $p^\prime$ is concentrated at small values of $|\theta|$, i.e., all higher moments $M_k$ are sufficiently small.
This means that for sufficiently concentrated and axially symmetric distributions, the shrinking factor $\mu$ is to a large extent solely determined by $\sigma$ and mostly independent from higher moments of $p^\prime$.
In particular, as shown in Fig.~\ref{fig:sigmaPlot}, in the range of $\sigma$ still allowing for a Bell violation, there is no significant difference in Bell parameters achieved by uniform patches and von Mises-Fisher distribution with the same $\sigma$.
In fact, for highly concentrated von Mises-Fisher distributions $\sigma \sim \kappa^{-1/2}$, justifying the scaling in Fig.~\ref{FIG_EXP_PATCHES} of the main text.
The failure of the quadratic approximation for larger distribution widths is indicated via comparison of the actual values of $\mu$ with expression~(\ref{eq:std_approx}), indicated by the black line in Fig.~\ref{fig:sigmaPlot}.

\subsection*{Trade-off between sharpness and statistical significance in acceptance-patch approach for uniform randomness}

\begin{figure}
    \centering
    \vspace{0.5cm}
    \includegraphics[width=0.9\linewidth]{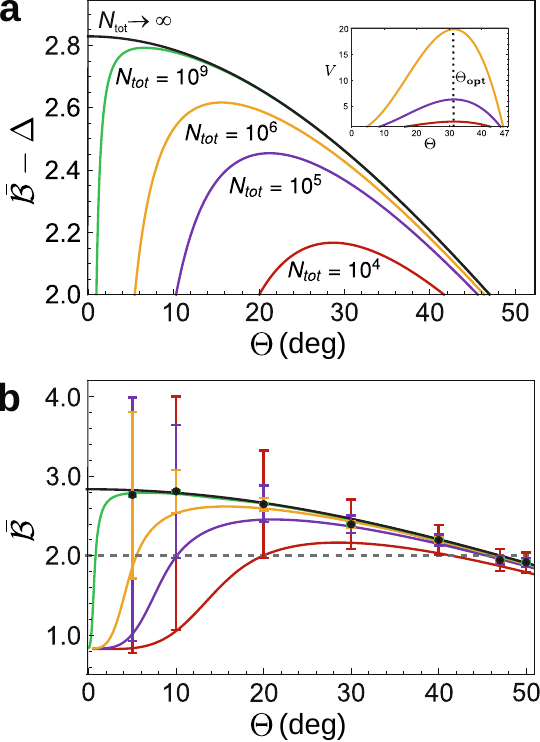} 
    \caption{Statistically significant violations of the CHSH inequality with uniformly random measurement directions as a function of the angular width of the patches centered at the optimal directions. $a)$ Theoretical predictions. For a finite number of measurements $N_{\mathrm{tot}}$, a statistically significant Bell violation, i.e., violation above at least one standard deviation $\Delta$ (colored curves), is obtained only within a limited range of patch widths. As shown on the inset, the optimal angular width, maximizing the violation in terms of the number of standard deviations $V$ [Eq.~(\ref{eq:V})], is independent of the total number of runs. $b)$ Experimental results. Each data point (black dots) corresponds to the patch-averaged CHSH parameter for a given angular width $\Theta$ and is obtained by processing the experimental data explained in the text. Coloured curves correspond to the ones plotted in panel $a)$ and agree with the estimated error shown through the colored error bars. 
}
    \label{FIG_LATY}
\end{figure}

As mentioned in the \textit{Results} section, there is a trade-off between the sharpness of the measurement which decreases with patch size, and the statistical significance of the violation, which increases with patch size (because a larger fraction of the Haar-randomly distributed directions fall into the patch).
Consider $N_{\mathrm{tot}}$ pairs of measurement directions are chosen uniformly at random and independently for Alice and Bob from \emph{entire Bloch spheres}.
Both observers set the centers of their Bloch sphere patches of width $\Theta$ at the optimal CHSH measurement directions.
Accordingly, the four correlation functions averaged over the respective patches for Alice and Bob are estimated using, on average,
$N_{\mathrm{tot}}\sin^4 (\Theta /2 )$ measurement runs.
The variance of each estimated correlation is $(1 - \bar E^2)  / N_{\mathrm{tot}}\sin^4 (\Theta /2 )$, where $\bar E$ is the averaged quantum mechanical prediction.
The Bell parameter itself is given by $\bar{\mathcal{B}} = \mu^2 \mathcal{B} = 2 \sqrt{2}\cos^4 (\Theta/2)$, where in the last equality we have used its optimal value and $\mu =\cos^2 (\Theta/2)$.
This optimal value is reached when the patches are centered at the optimal settings, in which case $\bar E = \mu^2/\sqrt{2}$.
Since four independently estimated correlations enter the Bell-CHSH parameter, the variance of the latter is multiplied by the factor of 4.
Given that, we estimate the error (standard deviation) of the CHSH value as
\begin{equation}
    \Delta=\sqrt{2} \, \left( \frac{2-\cos^8(\Theta/2)}{\sin^4(\Theta/2)} \frac{1}{N_{\mathrm{tot}}} \right)^{1/2}.
\end{equation}
The difference between the quantum value and the local bound in terms of the number of standard deviations is given by:
\begin{equation} \label{eq:V}
V =( 2 \sqrt{2} \cos^4(\Theta/2) - 2)/\Delta.
\end{equation}
Tab.~\ref{TAB_WINDOW} shows the minimal number of runs $N_{\mathrm{tot}}$ required to observe a violation by at least one standard deviation, i.e., $V \ge 1$, as well as the range of $\Theta$ where such violations occur. 
\begin{table}
\vspace{0.5cm}
\begin{tabular}{rccc}
\hline
$N_{\mathrm{tot}}$ & $\quad \Theta_{\min}$ & $\quad \Theta_{\max}$ & $\quad \Theta_{\mathrm{opt}} \quad$  \\
\hline
2,554 &  31.38$^{\circ}$ &  31.68$^{\circ}$  &31.53$^{\circ}$ \\
10,000 &  16.51$^{\circ}$ &  42.72$^{\circ}$ &  31.53$^{\circ}$  \\
100,000 &  8.64$^{\circ}$ &  45.83$^{\circ}$ &31.53$^{\circ}$  \\
1,000,000 &  4.77$^{\circ}$ &  46.65$^{\circ}$ &  31.53$^{\circ}$  \\
1,000,000,000 &  0.84$^{\circ}$ &  47.00$^{\circ}$ &  31.53$^{\circ}$  \\
\end{tabular}
\caption{\label{TAB_WINDOW} Conditions to violate CHSH inequality by at least one standard deviation, with uniformly sampled measurement directions and patches of angular width $\Theta$ centered at the optimal directions.
$N_{\mathrm{tot}}$ is the total number of experimental runs, $\Theta_{\min}$ stands for the minimal width of patches needed to observe the violation, $\Theta_{\max}$ for their maximal width, and $\Theta_{\mathrm{opt}}$ is the width of a patch that gives the largest statistically significant violation, i.e. violation by the highest number of standard deviations.
Note that in order to observe any meaningful violation, the experiment must be run at least $\approx 2500$ times. 
Note also that $\Theta_{\mathrm{opt}}$ is independent of $N_{\mathrm{tot}}$.
}
\end{table}
It turns out that the largest $V$ is achieved for $\Theta_{\mathrm{opt}} = 31.53^{\circ}$ independently of $N_{\mathrm{tot}}$.
Indeed, one finds that $\partial V / \partial \Theta = f(\Theta) \sqrt{N_{\mathrm{tot}}}$, and therefore the extremum condition $f(\Theta) = 0$ is independent of $N_{\mathrm{tot}}$. 

Fig.~\ref{FIG_LATY} shows explicitly how the value of the averaged Bell parameter depends on the angular width of the patch.
Note that even with a large $N_{\mathrm{tot}}$, say $10^9$, choosing a patch which is too small yields no violation as there are not enough data points falling within the patch to obtain a good approximation of the correlation function.  The theoretical predictions are accompanied by experimental data.
As each experiment performed for a fixed angular patch width $\Theta$ had $N_{\Theta}$ set to $10^4$, the data had to be processed by random sampling within sparser subsets.
For each patch width  $\Theta' \leq \Theta$ and each total number of measurements $N_{\mathrm{tot}}$, the experimental data were randomly partitioned into many effective realizations.
In each realization, the number of data points assigned to a given patch ($N_{\Theta'}$) was sampled from the corresponding binomial distribution, determined by the probability for a uniformly random measurement direction to fall within that patch.
The plotted values were obtained by averaging the resulting CHSH violations over all realizations.

\subsection*{Measurement directions for the icosahedral tiling}

 Table~\ref{tab:settings_ico} lists the labeled measurement directions used in the multi-setting Bell inequality (\ref{eq:multi_setting_CHSH}), optimized for the $|\phi^+\rangle$ state.
 These directions correspond to the normalized, tile-averaged Bloch vectors associated with the icosahedral tiling of the sphere.
 \begin{table}[!h]
 \caption{Measurement directions, represented as Bloch vectors, used in the multi-setting Bell inequality (\ref{eq:multi_setting_CHSH}) for the $|\phi^+\rangle$ state.}
 \centering
 \begin{tabular}{|c|c|}
 \hline
 1&$\{0, 0, 1\}$ \\ \hline
 2&$\{0, \frac{2}{3}, \frac{\sqrt{5}}{3}\}$ \\ \hline
 3&$\{-\frac{1}{\sqrt{3}}, -\frac{1}{3}, \frac{\sqrt{5}}{3}\}$ \\ \hline
 4&$\{\frac{1}{\sqrt{3}}, -\frac{1}{3}, \frac{\sqrt{5}}{3}\}$ \\ \hline
 5&$\{\frac{1}{\sqrt{3}}, \frac{\sqrt{5}}{3}, \frac{1}{3}\}$ \\ \hline
 6&$\{-\frac{1}{\sqrt{3}}, \frac{\sqrt{5}}{3}, \frac{1}{3}\}$ \\ \hline
 7&$\{\sqrt{\frac{1}{6}(3+\sqrt{5})}, \frac{1}{3}\sqrt{\frac{1}{2}(7-3\sqrt{5})}, \frac{1}{3}\}$ \\ \hline
 8&$\{-\sqrt{\frac{1}{6}(3+\sqrt{5})}, \frac{1}{3}\sqrt{\frac{1}{2}(7-3\sqrt{5})}, \frac{1}{3}\}$ \\ \hline
 9&$\{-\sqrt{\frac{1}{6}(3-\sqrt{5})}, -\sqrt{\frac{7}{18}+\frac{\sqrt{5}}{6}}, \frac{1}{3}\}$ \\ \hline
 10&$\{\sqrt{\frac{1}{6}(3-\sqrt{5})}, -\sqrt{\frac{7}{18}+\frac{\sqrt{5}}{6}}, \frac{1}{3}\}$ \\ \hline
 11&$\{\sqrt{\frac{1}{6}(3-\sqrt{5})}, \sqrt{\frac{7}{18}+\frac{\sqrt{5}}{6}}, -\frac{1}{3}\}$ \\ \hline
 12&$\{-\sqrt{\frac{1}{6}(3-\sqrt{5})}, \sqrt{\frac{7}{18}+\frac{\sqrt{5}}{6}}, -\frac{1}{3}\}$ \\ \hline
 13&$\{\sqrt{\frac{1}{6}(3+\sqrt{5})}, -\frac{1}{3}\sqrt{\frac{1}{2}(7-3\sqrt{5})}, -\frac{1}{3}\}$ \\ \hline
 14&$\{-\sqrt{\frac{1}{6}(3+\sqrt{5})}, -\frac{1}{3}\sqrt{\frac{1}{2}(7-3\sqrt{5})}, -\frac{1}{3}\}$ \\ \hline
 15&$\{-\frac{1}{\sqrt{3}}, -\frac{\sqrt{5}}{3}, -\frac{1}{3}\}$ \\ \hline
 16&$\{\frac{1}{\sqrt{3}}, -\frac{\sqrt{5}}{3}, -\frac{1}{3}\}$ \\ \hline
 17&$\{\frac{1}{\sqrt{3}}, \frac{1}{3}, -\frac{\sqrt{5}}{3}\}$ \\ \hline
 18&$\{-\frac{1}{\sqrt{3}}, \frac{1}{3}, -\frac{\sqrt{5}}{3}\}$ \\ \hline
 19&$\{0, -\frac{2}{3}, -\frac{\sqrt{5}}{3}\}$ \\ \hline
 20&$\{0, 0, -1\}$ \\ \hline
 \end{tabular}
 \label{tab:settings_ico}
 \end{table}

\subsection*{No Bell nonlocality from distributions of correlations}
\label{SEC_BELLENT}

Apart from the cases discussed in the main text, a further scenario of interest appears when the measurement directions for Alice and Bob are chosen randomly on the entire Bloch sphere, and outside of the control of the LHV, but the two parties do \textit{not} have access to information which settings have actually been chosen, such that neither acceptance-patches nor a Bloch-sphere tiling can be applied.
However, now for any pair of such generated directions, many copies of a state are measured.
Such an experiment has been realized, e.g., in~\cite{Knips2020}. 
Although it is not a usual Bell test, we ask whether the gathered data is sufficient to demonstrate Bell nonlocality.
In this scenario, the most informative quantity is the joint probability density $\chi(E_A, E_B, E_{AB})$, characterizing the likelihood that Alice (Bob) observes a local expectation value $E_A$ ($E_B$) and the correlation $E_{AB}$.
While moments of $\chi$ reveal quantum entanglement, even complete knowledge of $\chi$ cannot demonstrate Bell nonlocality because the statistics are intrinsically compatible with an LHV model.
Specifically, let the source sample $E_A, E_B, E_{AB}$ from $\chi$ and construct the joint probability
\begin{equation}
P(A,B) = \frac{1}{4} \left(1 + A \, E_A + B \, E_B + A B \, E_{AB} \right),
\label{EQ_RANDOM_LHV}
\end{equation}
from which it generates outcomes $A = \pm 1$ and $B = \pm 1$, sends them to Alice and Bob, whose devices output them independently of the actually chosen measurement directions.
This model reproduces the observed marginals and correlations and is fully local.
Note that all triples $(E_A, E_B, E_{AB})$ lead to $P(A,B) \geq 0$, since 
any physical distribution $\chi$ has to be compatible with
some quantum state $\rho$ which is positive semi-definite.
The model generalizes straightforwardly to more particles and higher dimensions.

One may also wonder whether limiting the measurements to a smaller range of directions, say a circle with a fixed $\theta$, could constrain the LHV models and allow to exclude them based on the joint distribution of correlation values.
Again, this is impossible.
Here, we construct an explicit LHV model for quantum correlations measured between observables parameterized only by local polar angles, $\theta_1$ for Alice and $\theta_2$ for Bob.
In this scenario, the correlations are averaged over local azimuthal angles, yielding
\begin{eqnarray}
    E(\theta_1, \theta_2)
    & = & \cos \theta_1 \cos \theta_2 T_{zz},
\end{eqnarray}
where $T_{\mu \nu} = \mathrm{tr}(\rho \sigma_{\mu} \otimes \sigma_{\nu})$, with $\sigma_0 = \mathbbm{1}$, such that $T_{zz}$ denotes the correlations of the underlying quantum state $\rho$ measured along the local $z$ axes.
Similarly, the local expectation values are
\begin{eqnarray}
    E(\theta_1) = \cos \theta_1 T_{z0} \quad \textrm{ and } \quad E(\theta_2) = \cos \theta_2 T_{0z}.
\end{eqnarray}
An LHV model which reproduces these statistics would be a source emitting hidden variables $\alpha$ and $\beta$ with probability
\begin{equation}
    P(\alpha, \beta) = \frac{1}{4} \left(1 + \alpha \, T_{z0} + \beta \, T_{0z} + \alpha \beta \, T_{zz} \right),
\end{equation}
analogous to the construction in the previous paragraph.
However, $\alpha = \pm 1$ and $\beta = \pm 1$ are not the measurement outcomes themselves.
Instead, the LHV model assigns measurement outcomes probabilistically, with the probabilities determined by the local measurement directions:
\begin{eqnarray}
    A & = & \alpha \quad \textrm{ with probability } \quad \frac{1}{2}(1 + \cos\theta_1) \\
    A & = & - \alpha \quad \textrm{ otherwise }
\end{eqnarray}
and similarly for Bob:
\begin{eqnarray}
    B & = & \beta \quad \textrm{ with probability } \quad \frac{1}{2}(1 + \cos\theta_2) \\
    B & = & - \beta \quad \textrm{ otherwise }
\end{eqnarray}
One verifies that this model reproduces the averaged quantum predictions for both the local expectation values and the correlation function.

\section*{Data availability}
The data that support the findings of this study are available from the corresponding author upon reasonable request.

\section*{Author Contributions}

 WL, TP, PC, and HW initialized the project. PC, JW, WL, TP, and PK developed the theory, which JD, CV, EMK, and HW later contributed to and refined. CV, EMK, JD, HW designed the experimental setup, performed the experiment, and evaluated the experimental data together with WL. All authors contributed to analyzing the results as well as writing the manuscript.

\section*{Funding}
PC is supported by the National Research Foundation, Singapore through the National Quantum Office, hosted in A*STAR, under its Centre for Quantum Technologies Funding Initiative (S24Q2d0009). WL, JW, PK, and TP acknowledge support of the National Science Centre (NCN, Poland) within the OPUS project (Grant No. 2024/53/B/ST2/04103). This work was funded by the Deutsche Forschungsgemeinschaft (German Research Foundation) under Germany’s Excellence Strategy – EXC-2111 – 390814868. EMK is supported by the education and training program of the Quantum Information Research Support Center, funded through the National research foundation of Korea (NRF) by the Ministry of science and ICT (MSIT) of the Korean government(No. RS-2023-NR057243).
We acknowledge the hospitality and support of the International Centre for Theory of Quantum Technologies (ICTQT), supported by project no. FENG.02.01-IP.05-0006/23, financed by the (IRAP) MAB FENG program 2021-2027, Priority FENG.02, Measure FENG.02.01., with the support of the FNP (Foundation for Polish Science).

\section*{Competing interests}
Authors declare no competing interests.

\bibliographystyle{quantum}
\bibliography{cite.bib}

\begin{thebibliography}{10}

\bibitem{Brunner2014}
Nicolas Brunner, Daniel Cavalcanti, Stefano Pironio, Valerio Scarani, and Stephanie Wehner.
\newblock ``Bell nonlocality''.
\newblock \href{https://dx.doi.org/10.1103/revmodphys.86.419}{Reviews of Modern Physics {\bf 86}, 419–478}~(2014).

\bibitem{Bell1964}
J.~S. Bell.
\newblock ``{On the Einstein Podolsky Rosen Paradox}''.
\newblock Physics {\bf 1}, 195--200~(1964).

\bibitem{Ekert1991}
Artur~K. Ekert.
\newblock ``{Quantum cryptography based on Bell’s theorem}''.
\newblock \href{https://dx.doi.org/10.1103/physrevlett.67.661}{Physical Review Letters {\bf 67}, 661–663}~(1991).

\bibitem{Mayers_1998}
D.~Mayers and A.~Yao.
\newblock ``Quantum cryptography with imperfect apparatus''.
\newblock In Proceedings 39th Annual Symposium on Foundations of Computer Science (Cat. No.98CB36280).
\newblock \href{https://dx.doi.org/10.1109/sfcs.1998.743501}{Page 503–509}.
\newblock SFCS-98. IEEE Comput. Soc~(1998).

\bibitem{Barrett_2005}
Jonathan Barrett, Lucien Hardy, and Adrian Kent.
\newblock ``No {S}ignaling and {Q}uantum {K}ey {D}istribution''.
\newblock \href{https://dx.doi.org/10.1103/PhysRevLett.95.010503}{Phys. Rev. Lett. {\bf 95}, 010503}~(2005).

\bibitem{Acin_2007}
Antonio Ac\'{\i}n, Nicolas Brunner, Nicolas Gisin, Serge Massar, Stefano Pironio, and Valerio Scarani.
\newblock ``Device-{I}ndependent {S}ecurity of {Q}uantum {C}ryptography against {C}ollective {A}ttacks''.
\newblock \href{https://dx.doi.org/10.1103/PhysRevLett.98.230501}{Phys. Rev. Lett. {\bf 98}, 230501}~(2007).

\bibitem{Pironio_2009}
Stefano Pironio, Antonio Acín, Nicolas Brunner, Nicolas Gisin, Serge Massar, and Valerio Scarani.
\newblock ``Device-independent quantum key distribution secure against collective attacks''.
\newblock \href{https://dx.doi.org/10.1088/1367-2630/11/4/045021}{New Journal of Physics {\bf 11}, 045021}~(2009).

\bibitem{Pironio_2010}
S.~Pironio, A.~Acín, S.~Massar, A.~Boyer de~la Giroday, D.~N. Matsukevich, P.~Maunz, S.~Olmschenk, D.~Hayes, L.~Luo, T.~A. Manning, and C.~Monroe.
\newblock ``Random numbers certified by {B}ell’s theorem''.
\newblock \href{https://dx.doi.org/10.1038/nature09008}{Nature {\bf 464}, 1021–1024}~(2010).

\bibitem{Colbeck_2011}
Roger Colbeck.
\newblock ``Quantum {A}nd {R}elativistic {P}rotocols {F}or {S}ecure {M}ulti-{P}arty {C}omputation''~(2011).
\newblock arXiv:0911.3814 [quant-ph].

\bibitem{Colbeck_2011_2}
Roger Colbeck and Adrian Kent.
\newblock ``Private randomness expansion with untrusted devices''.
\newblock \href{https://dx.doi.org/10.1088/1751-8113/44/9/095305}{Journal of Physics A: Mathematical and Theoretical {\bf 44}, 095305}~(2011).

\bibitem{ArnonFriedman_2018}
Rotem Arnon-Friedman, Frédéric Dupuis, Omar Fawzi, Renato Renner, and Thomas Vidick.
\newblock ``Practical device-independent quantum cryptography via entropy accumulation''.
\newblock Nature Communications {\bf 9}, 459~(2018).

\bibitem{Cleve_1997}
Richard Cleve and Harry Buhrman.
\newblock ``Substituting quantum entanglement for communication''.
\newblock \href{https://dx.doi.org/10.1103/PhysRevA.56.1201}{Phys. Rev. A {\bf 56}, 1201--1204}~(1997).

\bibitem{Brukner_2002}
{\v C}aslav Brukner, Marek \ifmmode~\dot{Z}\else \.{Z}\fi{}ukowski, and Anton Zeilinger.
\newblock ``Quantum {C}ommunication {C}omplexity {P}rotocol with {T}wo {E}ntangled {Q}utrits''.
\newblock \href{https://dx.doi.org/10.1103/PhysRevLett.89.197901}{Phys. Rev. Lett. {\bf 89}, 197901}~(2002).

\bibitem{Brukner_2004}
{\v C}aslav Brukner, Marek \ifmmode~\dot{Z}\else \.{Z}\fi{}ukowski, Jian-Wei Pan, and Anton Zeilinger.
\newblock ``Bell's inequalities and quantum communication complexity''.
\newblock \href{https://dx.doi.org/10.1103/PhysRevLett.92.127901}{Phys. Rev. Lett. {\bf 92}, 127901}~(2004).

\bibitem{Moreno_2020}
M.~G.~M. Moreno, Samura\'{\i} Brito, Ranieri~V. Nery, and Rafael Chaves.
\newblock ``Device-independent secret sharing and a stronger form of {B}ell nonlocality''.
\newblock \href{https://dx.doi.org/10.1103/PhysRevA.101.052339}{Phys. Rev. A {\bf 101}, 052339}~(2020).

\bibitem{Hensen2015}
B.~Hensen, H.~Bernien, A.~E. Dreau, A.~Reiserer, N.~Kalb, M.~S. Blok, J.~Ruitenberg, R.~F.~L. Vermeulen, R.~N. Schouten, C.~Abellan, W.~Amaya, V.~Pruneri, M.~W. Mitchell, M.~Markham, D.~J. Twitchen, D.~Elkouss, S.~Wehner, T.~H. Taminiau, and R.~Hanson.
\newblock ``{Loophole-free Bell inequality violation using electron spins separated by 1.3 kilometres}''.
\newblock \href{https://dx.doi.org/10.1038/nature15759}{Nature {\bf 526}, 682--686}~(2015).

\bibitem{Giustina2015}
M.~Giustina, M.~A.~M. Versteegh, S.~Wengerowsky, J.~Handsteiner, A.~Hochrainer, K.~Phelan, F.~Steinlechner, J.~Kofler, J.-A. Larsson, C.~Abell{\'a}n, W.~Amaya, V.~Pruneri, M.~W. Mitchell, J.~Beyer, T.~Gerrits, A.~E. Lita, L.~K. Shalm, S.~W. Nam, T.~Scheidl, R.~Ursin, B.~Wittmann, and A.~Zeilinger.
\newblock ``{Significant-Loophole-Free Test of Bell's Theorem with Entangled Photons}''.
\newblock \href{https://dx.doi.org/10.1103/PhysRevLett.115.250401}{Physical Review Letters {\bf 115}, 250401}~(2015).

\bibitem{Shalm2015}
L.~K. Shalm, E.~Meyer-Scott, B.~G. Christensen, P.~Bierhorst, M.~A. Wayne, M.~J. Stevens, T.~Gerrits, S.~Glancy, D.~R. Hamel, M.~S. Allman, K.~J. Coakley, S.~D. Dyer, C.~Hodge, A.~E. Lita, V.~B. Verma, C.~Lambrocco, E.~Tortorici, A.~L. Migdall, Y.~Zhang, W.~H. Farr, F.~Marsili, M.~D. Shaw, J.~Stern, C.~Abellan, W.~Amaya, V.~Pruneri, T.~Jennewein, M.~W. Mitchell, P.~G. Kwiat, R.~P. Mirin, E.~Knill, and S.~W. Nam.
\newblock ``{Strong Loophole-Free Test of Local Realism}''.
\newblock \href{https://dx.doi.org/10.1103/PhysRevLett.115.250402}{Physical Review Letters {\bf 115}, 250402}~(2015).

\bibitem{Rosenfeld2017}
Wenjamin Rosenfeld, Daniel Burchardt, Robert Garthoff, Kai Redeker, Norbert Ortegel, Markus Rau, and Harald Weinfurter.
\newblock ``{Event-Ready Bell Test Using Entangled Atoms Simultaneously Closing Detection and Locality Loopholes}''.
\newblock \href{https://dx.doi.org/10.1103/physrevlett.119.010402}{Physical Review Letters{\bf 119}}~(2017).

\bibitem{Rauch2018}
Dominik Rauch, Johannes Handsteiner, Jason Gallicchio, Andrew~S. Friedman, Bo~Li, Zhi-Peng Zhang, Johannes Kofler, David Bricher, Matthias Fink, Calvin Leung, Anthony Mark, Hien~T. Nguyen, Isabella Sanders, Fabian Steinlechner, Rupert Ursin, S{\"o}ren Wengerowsky, Thomas Scheidl, David~I. Kaiser, and Anton Zeilinger.
\newblock ``{Cosmic Bell Test Using Random Measurement Settings from High-Redshift Quasars}''.
\newblock \href{https://dx.doi.org/10.1103/PhysRevLett.121.080403}{Physical Review Letters {\bf 121}, 080403}~(2018).

\bibitem{Li_2018}
Ming-Han Li, Cheng Wu, Yanbao Zhang, Wen-Zhao Liu, Bing Bai, Yang Liu, Weijun Zhang, Qi~Zhao, Hao Li, Zhen Wang, Lixing You, W.~J. Munro, Juan Yin, Jun Zhang, Cheng-Zhi Peng, Xiongfeng Ma, Qiang Zhang, Jingyun Fan, and Jian-Wei Pan.
\newblock ``Test of local realism into the past without detection and locality loopholes''.
\newblock \href{https://dx.doi.org/10.1103/PhysRevLett.121.080404}{Phys. Rev. Lett. {\bf 121}, 080404}~(2018).

\bibitem{Storz_2023}
Simon Storz, Josua Sch{\"a}r, Anatoly Kulikov, Paul Magnard, Philipp Kurpiers, Janis L{\"u}tolf, Theo Walter, Adrian Copetudo, Kevin Reuer, Abdulkadir Akin, Jean-Claude Besse, Mihai Gabureac, Graham~J. Norris, Andr{\'e}s Rosario, Ferran Martin, Jos{\'e} Martinez, Waldimar Amaya, Morgan~W. Mitchell, Carlos Abellan, Jean-Daniel Bancal, Nicolas Sangouard, Baptiste Royer, Alexandre Blais, and Andreas Wallraff.
\newblock ``{Loophole-free Bell inequality violation with superconducting circuits}''.
\newblock \href{https://dx.doi.org/10.1038/s41586-023-05885-0}{Nature {\bf 617}, 265--270}~(2023).

\bibitem{Elben2022}
Andreas Elben, Steven~T. Flammia, Hsin-Yuan Huang, Richard Kueng, John Preskill, Benoit Vermersch, and Peter Zoller.
\newblock ``The randomized measurement toolbox''.
\newblock \href{https://dx.doi.org/10.1038/s42254-022-00535-2}{Nature Reviews Physics {\bf 5}, 9–24}~(2022).

\bibitem{Cieliski2024}
Paweł Cieśliński, Satoya Imai, Jan Dziewior, Otfried G\"{u}hne, Lukas Knips, Wiesław Laskowski, Jasmin Meinecke, Tomasz Paterek, and Tamás Vértesi.
\newblock ``Analysing quantum systems with randomised measurements''.
\newblock \href{https://dx.doi.org/10.1016/j.physrep.2024.09.009}{Physics Reports {\bf 1095}, 1–48}~(2024).

\bibitem{Huang2020}
Hsin-Yuan Huang, Richard Kueng, and John Preskill.
\newblock ``Predicting many properties of a quantum system from very few measurements''.
\newblock \href{https://dx.doi.org/10.1038/s41567-020-0932-7}{Nature Physics {\bf 16}, 1050–1057}~(2020).

\bibitem{Vermersch2024}
Benoit Vermersch, Aniket Rath, Bharathan Sundar, Cyril Branciard, John Preskill, and Andreas Elben.
\newblock ``Enhanced estimation of quantum properties with common randomized measurements''.
\newblock \href{https://dx.doi.org/10.1103/prxquantum.5.010352}{PRX Quantum {\bf 5}, 010352}~(2024).

\bibitem{Brun2004}
Todd Brun.
\newblock ``Measuring polynomial functions of states''.
\newblock \href{https://dx.doi.org/}{Quantum Inf. Comp. {\bf 4}, 401--408}~(2004).

\bibitem{vanEnk2012}
S.~J. van Enk and C.~W.~J. Beenakker.
\newblock ``Measuring $\mathrm{Tr}(\rho^n)$ on single copies of $\rho$ using random measurements''.
\newblock \href{https://dx.doi.org/10.1103/physrevlett.108.110503}{Physical Review Letters {\bf 108}, 110503}~(2012).

\bibitem{Knips2020}
Lukas Knips, Jan Dziewior, Waldemar Kłobus, Wiesław Laskowski, Tomasz Paterek, Peter~J. Shadbolt, Harald Weinfurter, and Jasmin D.~A. Meinecke.
\newblock ``Multipartite entanglement analysis from random correlations''.
\newblock \href{https://dx.doi.org/10.1038/s41534-020-0281-5}{npj Quantum Information {\bf 6}, 51}~(2020).

\bibitem{Elben2020topo}
Andreas Elben, Jinlong Yu, Guanyu Zhu, Mohammad Hafezi, Frank Pollmann, Peter Zoller, and Benoit Vermersch.
\newblock ``Many-body topological invariants from randomized measurements in synthetic quantum matter''.
\newblock \href{https://dx.doi.org/10.1126/sciadv.aaz3666}{Science Advances {\bf 6}, aaz3666}~(2020).

\bibitem{Tran2015}
Minh~Cong Tran, Borivoje Dakić, Fran\c{c}ois Arnault, Wiesław Laskowski, and Tomasz Paterek.
\newblock ``Quantum entanglement from random measurements''.
\newblock \href{https://dx.doi.org/10.1103/physreva.92.050301}{Physical Review A {\bf 92}, 050301}~(2015).

\bibitem{Tran2016}
Minh~Cong Tran, Borivoje Dakić, Wiesław Laskowski, and Tomasz Paterek.
\newblock ``Correlations between outcomes of random measurements''.
\newblock \href{https://dx.doi.org/10.1103/physreva.94.042302}{Physical Review A {\bf 94}, 042302}~(2016).

\bibitem{Brydges2019}
Tiff Brydges, Andreas Elben, Petar Jurcevic, Benoit Vermersch, Christine Maier, Ben~P. Lanyon, Peter Zoller, Rainer Blatt, and Christian~F. Roos.
\newblock ``{Probing Rényi entanglement entropy via randomized measurements}''.
\newblock \href{https://dx.doi.org/10.1126/science.aau4963}{Science {\bf 364}, 260–263}~(2019).

\bibitem{Ketterer2019}
Andreas Ketterer, Nikolai Wyderka, and Otfried G\"{u}hne.
\newblock ``Characterizing multipartite entanglement with moments of random correlations''.
\newblock \href{https://dx.doi.org/10.1103/physrevlett.122.120505}{Physical Review Letters {\bf 122}, 120505}~(2019).

\bibitem{Wyderka2020}
N~Wyderka and O~G\"{u}hne.
\newblock ``Characterizing quantum states via sector lengths''.
\newblock \href{https://dx.doi.org/10.1088/1751-8121/ab7f0a}{Journal of Physics A: Mathematical and Theoretical {\bf 53}, 345302}~(2020).

\bibitem{Zhou2020}
You Zhou, Pei Zeng, and Zhenhuan Liu.
\newblock ``Single-copies estimation of entanglement negativity''.
\newblock \href{https://dx.doi.org/10.1103/physrevlett.125.200502}{Physical Review Letters {\bf 125}, 200502}~(2020).

\bibitem{Elben2020}
Andreas Elben, Richard Kueng, Hsin-Yuan~(Robert) Huang, Rick van Bijnen, Christian Kokail, Marcello Dalmonte, Pasquale Calabrese, Barbara Kraus, John Preskill, Peter Zoller, and Benoit Vermersch.
\newblock ``Mixed-state entanglement from local randomized measurements''.
\newblock \href{https://dx.doi.org/10.1103/physrevlett.125.200501}{Physical Review Letters {\bf 125}, 200501}~(2020).

\bibitem{Yu2021}
Xiao-Dong Yu, Satoya Imai, and Otfried G\"{u}hne.
\newblock ``Optimal entanglement certification from moments of the partial transpose''.
\newblock \href{https://dx.doi.org/10.1103/physrevlett.127.060504}{Physical Review Letters {\bf 127}, 060504}~(2021).

\bibitem{Imai_2021}
Satoya Imai, Nikolai Wyderka, Andreas Ketterer, and Otfried G\"uhne.
\newblock ``Bound entanglement from randomized measurements''.
\newblock \href{https://dx.doi.org/10.1103/PhysRevLett.126.150501}{Phys. Rev. Lett. {\bf 126}, 150501}~(2021).

\bibitem{Neven2021}
Antoine Neven, Jose Carrasco, Vittorio Vitale, Christian Kokail, Andreas Elben, Marcello Dalmonte, Pasquale Calabrese, Peter Zoller, Benoit Vermersch, Richard Kueng, and Barbara Kraus.
\newblock ``Symmetry-resolved entanglement detection using partial transpose moments''.
\newblock \href{https://dx.doi.org/10.1038/s41534-021-00487-y}{npj Quantum Information {\bf 7}, 152}~(2021).

\bibitem{Ketterer_2022}
Andreas Ketterer, Satoya Imai, Nikolai Wyderka, and Otfried G\"uhne.
\newblock ``Statistically significant tests of multiparticle quantum correlations based on randomized measurements''.
\newblock \href{https://dx.doi.org/10.1103/PhysRevA.106.L010402}{Phys. Rev. A {\bf 106}, L010402}~(2022).

\bibitem{Liang_2010}
Yeong-Cherng Liang, Nicholas Harrigan, Stephen~D. Bartlett, and Terry Rudolph.
\newblock ``Nonclassical {C}orrelations from {R}andomly {C}hosen {L}ocal {M}easurements''.
\newblock \href{https://dx.doi.org/10.1103/PhysRevLett.104.050401}{Phys. Rev. Lett. {\bf 104}, 050401}~(2010).

\bibitem{Shadbolt_2012}
Peter Shadbolt, Tamás Vértesi, Yeong-Cherng Liang, Cyril Branciard, Nicolas Brunner, and Jeremy~L. O’Brien.
\newblock ``Guaranteed violation of a {B}ell inequality without aligned reference frames or calibrated devices''.
\newblock \href{https://dx.doi.org/10.1038/srep00470}{Scientific Reports{\bf 2}}~(2012).

\bibitem{Wallman_2012}
Joel~J. Wallman and Stephen~D. Bartlett.
\newblock ``Observers can always generate nonlocal correlations without aligning measurements by covering all their bases''.
\newblock \href{https://dx.doi.org/10.1103/PhysRevA.85.024101}{Phys. Rev. A {\bf 85}, 024101}~(2012).

\bibitem{Fonseca_2015}
E.~A. Fonseca and Fernando Parisio.
\newblock ``Measure of nonlocality which is maximal for maximally entangled qutrits''.
\newblock \href{https://dx.doi.org/10.1103/PhysRevA.92.030101}{Phys. Rev. A {\bf 92}, 030101(R)}~(2015).

\bibitem{Senel_2015}
Celal Furkan~Senel, Thomas Lawson, Marc Kaplan, Damian Markham, and Eleni Diamanti.
\newblock ``Demonstrating genuine multipartite entanglement and nonseparability without shared reference frames''.
\newblock \href{https://dx.doi.org/10.1103/PhysRevA.91.052118}{Phys. Rev. A {\bf 91}, 052118}~(2015).

\bibitem{de_Rosier_2017}
Anna de~Rosier, Jacek Gruca, Fernando Parisio, Tam\'as V\'ertesi, and Wies\l{}aw Laskowski.
\newblock ``Multipartite nonlocality and random measurements''.
\newblock \href{https://dx.doi.org/10.1103/PhysRevA.96.012101}{Phys. Rev. A {\bf 96}, 012101}~(2017).

\bibitem{Lipinska_2018}
Victoria Lipinska, Florian~J. Curchod, Alejandro Máttar, and Antonio Acín.
\newblock ``Towards an equivalence between maximal entanglement and maximal quantum nonlocality''.
\newblock \href{https://dx.doi.org/10.1088/1367-2630/aaca22}{New Journal of Physics {\bf 20}, 063043}~(2018).

\bibitem{Barasinski_2019}
Artur Barasi\ifmmode~\acute{n}\else \'{n}\fi{}ski, Anton\'{\i}n \ifmmode~\check{C}\else \v{C}\fi{}ernoch, Karel Lemr, and Jan Soubusta.
\newblock ``Experimental verification of time-order-dependent correlations in three-qubit {G}reenberger-{H}orne-{Z}eilinger-class states''.
\newblock \href{https://dx.doi.org/10.1103/PhysRevA.99.042123}{Phys. Rev. A {\bf 99}, 042123}~(2019).

\bibitem{Barasinski_2020}
Artur Barasi\ifmmode~\acute{n}\else \'{n}\fi{}ski, Anton\'{\i}n \ifmmode~\check{C}\else \v{C}\fi{}ernoch, Karel Lemr, and Jan Soubusta.
\newblock ``Genuine tripartite nonlocality for random measurements in {G}reenberger-{H}orne-{Z}eilinger-class states and its experimental test''.
\newblock \href{https://dx.doi.org/10.1103/PhysRevA.101.052109}{Phys. Rev. A {\bf 101}, 052109}~(2020).

\bibitem{Yang_2020}
Shih-Xian Yang, Gelo~Noel Tabia, Pei-Sheng Lin, and Yeong-Cherng Liang.
\newblock ``Device-independent certification of multipartite entanglement using measurements performed in randomly chosen triads''.
\newblock \href{https://dx.doi.org/10.1103/PhysRevA.102.022419}{Phys. Rev. A {\bf 102}, 022419}~(2020).

\bibitem{de_Rosier_2020}
Anna de~Rosier, Jacek Gruca, Fernando Parisio, Tam\'as V\'ertesi, and Wies\l{}aw Laskowski.
\newblock ``Strength and typicality of nonlocality in multisetting and multipartite {B}ell scenarios''.
\newblock \href{https://dx.doi.org/10.1103/PhysRevA.101.012116}{Phys. Rev. A {\bf 101}, 012116}~(2020).

\bibitem{Pandit_2022}
Mahasweta Pandit, Artur Barasiński, István Márton, Tamás Vértesi, and Wiesław Laskowski.
\newblock ``Optimal tests of genuine multipartite nonlocality''.
\newblock \href{https://dx.doi.org/10.1088/1367-2630/aca8c8}{New Journal of Physics {\bf 24}, 123017}~(2022).

\bibitem{Barasinski_2021}
Artur Barasi{\'{n}}ski, Anton{\'{i}}n {\v{C}}ernoch, Wies{\l{}}aw Laskowski, Karel Lemr, Tam{\'{a}}s V{\'{e}}rtesi, and Jan Soubusta.
\newblock ``{Experimentally friendly approach towards nonlocal correlations in multisetting {N}-partite {B}ell scenarios}''.
\newblock \href{https://dx.doi.org/10.22331/q-2021-04-14-430}{{Quantum} {\bf 5}, 430}~(2021).

\bibitem{Cieslinski_2025}
Paweł Cieśliński, Tamás Vértesi, Mateusz Kowalczyk, and Wiesław Laskowski.
\newblock ``How likely are you to observe non-locality with imperfect detection efficiency and random measurement settings?''.
\newblock \href{https://dx.doi.org/10.1088/1367-2630/ade90f}{New Journal of Physics {\bf 27}, 074504}~(2025).

\bibitem{Clauser1969}
John~F. Clauser, Michael~A. Horne, Abner Shimony, and Richard~A. Holt.
\newblock ``Proposed experiment to test local hidden-variable theories''.
\newblock \href{https://dx.doi.org/10.1103/physrevlett.23.880}{Physical Review Letters {\bf 23}, 880–884}~(1969).

\bibitem{Busch1986}
Paul Busch.
\newblock ``Unsharp reality and joint measurements for spin observables''.
\newblock \href{https://dx.doi.org/10.1103/physrevd.33.2253}{Physical Review D {\bf 33}, 2253–2261}~(1986).

\bibitem{Fine1982}
Arthur Fine.
\newblock ``{Hidden Variables, Joint Probability, and the Bell Inequalities}''.
\newblock \href{https://dx.doi.org/10.1103/physrevlett.48.291}{Physical Review Letters {\bf 48}, 291–295}~(1982).

\bibitem{ClauserShimonyReview}
J~F Clauser and A~Shimony.
\newblock ``Bell’s theorem. experimental tests and implications''.
\newblock \href{https://dx.doi.org/10.1088/0034-4885/41/12/002}{Reports on Progress in Physics {\bf 41}, 1881–1927}~(1978).

\bibitem{Larsson2014}
Jan-$\AA$ke Larsson.
\newblock ``{Loopholes in Bell inequality tests of local realism}''.
\newblock \href{https://dx.doi.org/10.1088/1751-8113/47/42/424003}{Journal of Physics A: Mathematical and Theoretical {\bf 47}, 424003}~(2014).

\bibitem{Hoi_2025}
Hoi~Ping Luk, Roman Nedela, and Christopher Purcell.
\newblock ``Tiling the sphere with regular polygons''~(2025).

\bibitem{Johnson1966}
Norman~W. Johnson.
\newblock ``Convex polyhedra with regular faces''.
\newblock \href{https://dx.doi.org/10.4153/cjm-1966-021-8}{Canadian Journal of Mathematics {\bf 18}, 169–200}~(1966).

\bibitem{zalgaller1969convex}
V.~A. Zalgaller.
\newblock ``Convex polyhedra with regular faces''.
\newblock Volume~2 of Seminars in Mathematics, V. A. Steklov Mathematical Institute, Leningrad.
\newblock Consultants Bureau. New York~(1969).

\bibitem{Evans2010}
P.~G. Evans, R.~S. Bennink, W.~P. Grice, T.~S. Humble, and J.~Schaake.
\newblock ``Bright source of spectrally uncorrelated polarization-entangled photons with nearly single-mode emission''.
\newblock \href{https://dx.doi.org/10.1103/PhysRevLett.105.253601}{Physical Review Letters {\bf 105}, 253601}~(2010).

\bibitem{Werner1989}
Reinhard~F. Werner.
\newblock ``{Quantum states with Einstein-Podolsky-Rosen correlations admitting a hidden-variable model}''.
\newblock \href{https://dx.doi.org/10.1103/physreva.40.4277}{Physical Review A {\bf 40}, 4277–4281}~(1989).

\bibitem{Augusiak2015}
R.~Augusiak, M.~Demianowicz, J.~Tura, and A.~Acín.
\newblock ``Entanglement and nonlocality are inequivalent for any number of parties''.
\newblock \href{https://dx.doi.org/10.1103/physrevlett.115.030404}{Physical Review Letters {\bf 115}, 030404}~(2015).

\bibitem{Quintino2015}
Marco~Túlio Quintino, Tamás Vértesi, Daniel Cavalcanti, Remigiusz Augusiak, Maciej Demianowicz, Antonio Acín, and Nicolas Brunner.
\newblock ``{Inequivalence of entanglement, steering, and Bell nonlocality for general measurements}''.
\newblock \href{https://dx.doi.org/10.1103/physreva.92.032107}{Physical Review A {\bf 92}, 032107}~(2015).

\bibitem{Anwer2021}
Hammad Anwer, Natalie Wilson, Ralph Silva, Sadiq Muhammad, Armin Tavakoli, and Mohamed Bourennane.
\newblock ``Noise-robust preparation contextuality shared between any number of observers via unsharp measurements''.
\newblock \href{https://dx.doi.org/10.22331/q-2021-09-28-551}{Quantum {\bf 5}, 551}~(2021).

\bibitem{Miklin_2020}
Nikolai Miklin, Jakub~J. Borka\l{}a, and Marcin Paw\l{}owski.
\newblock ``Semi-device-independent self-testing of unsharp measurements''.
\newblock \href{https://dx.doi.org/10.1103/PhysRevResearch.2.033014}{Phys. Rev. Res. {\bf 2}, 033014}~(2020).

\bibitem{Anwer_2020}
Hammad Anwer, Sadiq Muhammad, Walid Cherifi, Nikolai Miklin, Armin Tavakoli, and Mohamed Bourennane.
\newblock ``Experimental characterization of unsharp qubit observables and sequential measurement incompatibility via quantum random access codes''.
\newblock \href{https://dx.doi.org/10.1103/PhysRevLett.125.080403}{Phys. Rev. Lett. {\bf 125}, 080403}~(2020).

\bibitem{Curchod_2017}
F.~J. Curchod, M.~Johansson, R.~Augusiak, M.~J. Hoban, P.~Wittek, and A.~Ac\'{\i}n.
\newblock ``Unbounded randomness certification using sequences of measurements''.
\newblock \href{https://dx.doi.org/10.1103/PhysRevA.95.020102}{Phys. Rev. A {\bf 95}, 020102}~(2017).

\bibitem{Silva_2015}
Ralph Silva, Nicolas Gisin, Yelena Guryanova, and Sandu Popescu.
\newblock ``Multiple observers can share the nonlocality of half of an entangled pair by using optimal weak measurements''.
\newblock \href{https://dx.doi.org/10.1103/PhysRevLett.114.250401}{Phys. Rev. Lett. {\bf 114}, 250401}~(2015).

\bibitem{Palacios-Laloy_2010}
Agustin Palacios-Laloy, Fran\c{c}ois Mallet, Fran\c{c}ois Nguyen, Patrice Bertet, Denis Vion, Daniel Esteve, and Alexander~N. Korotkov.
\newblock ``{Experimental violation of a Bell’s inequality in time with weak measurement}''.
\newblock \href{https://dx.doi.org/10.1038/nphys1641}{Nature Physics {\bf 6}, 442–447}~(2010).

\bibitem{Armen_2002}
Michael~A. Armen, John~K. Au, John~K. Stockton, Andrew~C. Doherty, and Hideo Mabuchi.
\newblock ``Adaptive homodyne measurement of optical phase''.
\newblock \href{https://dx.doi.org/10.1103/PhysRevLett.89.133602}{Phys. Rev. Lett. {\bf 89}, 133602}~(2002).

\bibitem{Audretsch_2001}
J\"urgen Audretsch, Thomas Konrad, and Artur Scherer.
\newblock ``Sequence of unsharp measurements enabling a real-time visualization of a quantum oscillation''.
\newblock \href{https://dx.doi.org/10.1103/PhysRevA.63.052102}{Phys. Rev. A {\bf 63}, 052102}~(2001).

\bibitem{Bartlett_2006}
Stephen~D Bartlett, Terry Rudolph, Robert~W Spekkens, and Peter~S Turner.
\newblock ``Degradation of a quantum reference frame''.
\newblock \href{https://dx.doi.org/10.1088/1367-2630/8/4/058}{New Journal of Physics {\bf 8}, 58–58}~(2006).

\bibitem{Vaccaro_2008}
J.~A. Vaccaro, F.~Anselmi, H.~M. Wiseman, and K.~Jacobs.
\newblock ``Tradeoff between extractable mechanical work, accessible entanglement, and ability to act as a reference system, under arbitrary superselection rules''.
\newblock \href{https://dx.doi.org/10.1103/PhysRevA.77.032114}{Phys. Rev. A {\bf 77}, 032114}~(2008).

\bibitem{Gour_2008}
Gilad Gour and Robert~W Spekkens.
\newblock ``The resource theory of quantum reference frames: manipulations and monotones''.
\newblock \href{https://dx.doi.org/10.1088/1367-2630/10/3/033023}{New Journal of Physics {\bf 10}, 033023}~(2008).

\bibitem{Gour_2009}
Gilad Gour, Iman Marvian, and Robert~W. Spekkens.
\newblock ``Measuring the quality of a quantum reference frame: The relative entropy of frameness''.
\newblock \href{https://dx.doi.org/10.1103/PhysRevA.80.012307}{Phys. Rev. A {\bf 80}, 012307}~(2009).

\bibitem{Barlett_2007}
Stephen~D. Bartlett, Terry Rudolph, and Robert~W. Spekkens.
\newblock ``Reference frames, superselection rules, and quantum information''.
\newblock \href{https://dx.doi.org/10.1103/RevModPhys.79.555}{Rev. Mod. Phys. {\bf 79}, 555--609}~(2007).

\bibitem{Kofler2007}
Johannes Kofler and Časlav Brukner.
\newblock ``Classical world arising out of quantum physics under the restriction of coarse-grained measurements''.
\newblock \href{https://dx.doi.org/10.1103/physrevlett.99.180403}{Physical Review Letters {\bf 99}, 180403}~(2007).

\bibitem{Kofler2008}
Johannes Kofler and Časlav Brukner.
\newblock ``Conditions for quantum violation of macroscopic realism''.
\newblock \href{https://dx.doi.org/10.1103/physrevlett.101.090403}{Physical Review Letters {\bf 101}, 090403}~(2008).

\bibitem{Teh2023}
Run~Yan Teh, Laura Rosales-Zarate, Peter~D. Drummond, and M.D. Reid.
\newblock ``Mesoscopic and macroscopic quantum correlations in photonic, atomic and optomechanical systems''.
\newblock \href{https://dx.doi.org/10.1016/j.pquantelec.2022.100396}{Progress in Quantum Electronics {\bf 90}, 100396}~(2023).

\bibitem{Leggett1985}
A.~J. Leggett and Anupam Garg.
\newblock ``Quantum mechanics versus macroscopic realism: Is the flux there when nobody looks?''.
\newblock \href{https://dx.doi.org/10.1103/physrevlett.54.857}{Physical Review Letters {\bf 54}, 857–860}~(1985).

\end{thebibliography}

\end{document}